 \documentclass[preprint]{aastex}
\usepackage{psfig}
\usepackage{longtable}

\begin{document}

\title{Temperature Map 
	of the Virgo Cluster of Galaxies 
	Observed with ASCA}

\author{
R. Shibata                 \altaffilmark{1},
K. Matsushita              \altaffilmark{2},
N. Y. Yamasaki             \altaffilmark{3},
T. Ohashi                  \altaffilmark{3},
\\
M. Ishida                  \altaffilmark{1},
K. Kikuchi                 \altaffilmark{4},
H. B\"ohringer             \altaffilmark{2},
and 
H. Matsumoto               \altaffilmark{5}
}

\altaffiltext{1}{Institute of Space and Astronautical Science, 3-1-1, 
Yoshinodai,Sagamihara, Kanagawa 229-8510 Japan}

\altaffiltext{2}{Max-Planck-Institut f$\ddot{\rm u}$r 
extraterrestrische Physik,
Giessenbachstrasse Postfach 1603, Garching, D-85740, Germany}

\altaffiltext{3}{Department of Physics, Tokyo Metropolitan University, 
1-1 Minami-Osawa, Hachioji, Tokyo 192-0397 Japan}

\altaffiltext{4}{Space Utilization Research Program, 
National Space Development 
Agency of Japan, 2-1-1 Sengen, Tshukuba, Ibaraki 305-8505, Japan}

\altaffiltext{5}{Center for Space Research, 
Massachusetts Institute of Technology,
77 Massachusetts Avenue, Cambridge, MA 02139-4307, USA}

\authoremail{shibata@astro.isas.ac.jp}

\begin{abstract}

The temperature distribution of the intracluster medium (ICM) in the
Virgo cluster of galaxies has been derived from extensive mapping
observations with ASCA covering an area of 19 deg$^2$. In the spectral
analysis, the inner region within a radius of $\sim 60'$ from M87 is
characterized by an ICM temperature of $kT \sim 2.5$ keV\@ with
little variation. On the other hand, the outer regions indicate
significant variation of the temperature with an amplitude of about 1
keV\@. The temperature map was produced from the hardness ratio (H.R.) 
values with a resolution of about $5'$. Besides the previously
reported hot region with $kT > 4$ keV between M87 and M49, several hot
regions with $kT = 3-4$ keV are detected in the cluster outskirts.
The auto-correlation function for the H.R. variation shows that the
temperature variation is correlated within a size of about 300 kpc,
suggesting that gas blobs falling in the Virgo cluster have a typical
size of groups of galaxies.  A correlation with the velocity dispersion
of member galaxies shows that only the north-west region indicates an
unusually large $\beta_{\rm spec}$ value of $2-4$. The upper limit for
extended non-thermal emission in the Virgo cluster is obtained to
be $L_X \sim 9 \times 10^{41}$ ergs s$^{-1}$ in the 2 $-$ 10 keV band. 
We discuss that these features consistently indicate that the Virgo
cluster is in a relatively early stage of the cluster evolution.

\end{abstract}


\section{Introduction}

Clusters of galaxies are believed to be evolving over cosmological
time scales. Studies of the redshift evolution of the X-ray luminosity
function of clusters have been carried out, and the observed results
indicate that the most luminous clusters tend to be sparse in the
redshift range greater than about 0.3 (Rosati et al.\ 1998). Recently,
evidence for the evolution of individual clusters have been obtained
in the form of temperature variations in the intracluster medium (ICM)
from more than 10 systems (Henrikson \& Markevitch 1996; Henry \&
Briel 1996; Honda, et al.\ 1996; Davis \& White 1998; Donnelly et
al. 1998; Markevitch et al.\ 1998; Churazov et al.\ 1999; Markevitch et
al. 1999; Molendi et al.\ 1999; Shibata et al.\ 1999; Watanabe et
al. 1999). It is considered that gas heating is caused by collisions
between subclusters and main clusters, and the temperatures are raised
by several keV\@. Detailed mappings of the temperature distribution in
nearby clusters would give us information about the size and number
distribution of the colliding blobs, which can be used to constrain
theoretical models of structure formation.

The Virgo cluster of galaxies is the nearest cluster, with the
enormously large angular extent of about $6^{\circ}$ in radius. It is
classified as Bautz-Morgan type III (Bautz \& Morgan 1970), and
Rood-Sastry type F (Rood \& Sastry 1971) and shows a highly irregular
morphology (Abell, Corwin, \& Olowin 1989). This feature suggests that
the cluster is undergoing dynamical evolution.

In the optical band, a catalogue containing 2096 galaxies within an
area of $\sim 140$ deg$^2$ roughly centered at $\alpha \sim 12^{\rm h}
15^{\rm m}$ and $\delta \sim 13^{\circ}$ was produced by Binggeli,
Sandage, \& Tammann (1985) and Binggeli \& Cameron (1993).  Optical
properties of about 1300 member galaxies, such as morphologies, number
counts, and heliocentric velocities, have been examined.  The distance
to the cluster was measured with various methods and derived to be $12
- 22$ Mpc with a large uncertainty (summarized in Jacoby et al.\
1990).  Recently, Federspiel, Tammann, \& Sandage (1998) derived $20.7
\pm 2.4$ Mpc based on the 21 cm line width-absolute magnitude relation
in the $B$ band.  Fukugita, Okamura, \& Yasuda (1993) and Yasuda,
Fukugita, \& Okamura (1997), based on the Tully-Fisher relation for
spiral galaxies, showed evidence that the galaxy distribution is
largely elongated in the line of sight from 12 to 30 Mpc.  Jacoby et
al.\ (1990) showed that the M49 and M86 subclusters were located $\sim
0.6 \pm 0.7$ Mpc and $1.2 \pm 0.6$ Mpc farther than the M87 subcluster
on average.  The M86 distance is consistent with the result by Neilsen
\& Tsvetanov (2000, $2.5 \pm 1.5$ Mpc).

In the X-ray band, the first indication of very extended emission over
angular scales of $10^{\circ}$ was obtained with non-imaging detectors
on board the Ginga satellite (Takano 1990; Koyama, Takano, \& Tawara
1991).  The X-ray emission is strongly peaked at M87, and well
described by a $\beta$ model with $\beta = 0.436$, $R_{\rm c} =
1'.62$, as previously obtained by the Einstein IPC data within 100$'$
from M87 (Fabricant \& Gorenstein 1983).  In the southern outer part
(2$^{\circ}$ away from M87), the X-ray surface brightness distribution
shows significant flattening with $\beta \sim 0.357$.  The ICM
temperature around M87 was almost constant with $kT \sim 2$ keV, but
the Ginga data suggested a temperature rise toward M49\@.

The ROSAT all-sky survey covered the Virgo cluster region between November
1990 and January 1991 in the energy range $0.1 - 2.4$ keV
(B\"{o}hringer et al.\ 1994).  On large scales, the X-ray
morphology generally traces the galaxy distribution (Schindler,
Binggeli, \& B\"{o}hringer 1999).  However, the M49 halo is much
smaller in the X-ray image compared to the optical appearance of the
galaxies around M49\@.  The X-ray surface brightness around M87 can be
fitted by the $\beta$ model with $\beta = 0.45$ and $R_{\rm c} =
2'.3$.  Although the overall temperature distribution around M87 is
consistent with the Ginga result within $\sim 1^{\circ}$, ROSAT showed
no evidence for the temperature rise around M49\@.  Recently, mapping
observations from ASCA have detected an unusual hot region between M87
and M49 with a temperature twice as high as the surrounding value
of 2 keV (Kikuchi et al.\ 2000). This provides clear evidence that
the Virgo cluster is undergoing strong dynamical evolution.

In this paper, we have significantly expanded the mapped area and
present a nearly complete view of the brightness and temperature
distributions of the ICM\@.  The closeness of this cluster enables us
to look into detailed structures in the temperature distribution at a
resolution of a few hundred kpc for the first time.  Results on the
metal distribution will be reported in a separate paper (Shibata et
al.\ 2000).  We assume the distance to be 20 Mpc (e.g., Federspiel,
Tammann, \& Sandage 1998), and a $1'$ angular separation corresponds
to 5.8 kpc.  The solar number abundance of Fe relative to H is taken
as $4.68 \times 10^{-5}$ (Anders \& Grevesse 1989) throughout this
paper.

\section{Observation and Analysis}
\subsection{ASCA Observations}

The ASCA observations of the Virgo cluster have been carried out in 3
occasions: December 1996 and June 1997 with a combined exposure of 200
ksec (northwest region of M87, Matsumoto et al.\ 2000), June 1998 for
280 ksec (linking region between M87 and M49, Kikuchi et al.\ 2000),
and June 1999 for 210 ksec (region around M87), respectively.  There
are other 19 pointings on known X-ray sources whose locations are
within the Virgo system. Including these data, the total number of
pointings is 60 and the covered sky area is $\sim 19$ deg$^2$,
respectively. Figure 1 shows the regions so far observed with ASCA\@.
The radius of each circle is $22'$ corresponding to the field of view
(FOV) of the Gas Imaging Spectrometer (GIS; Ohashi et al.\ 1996;
Makishima et al.\ 1996).  The observation log is shown in table 1.

The ASCA satellite (Tanaka, Inoue, \& Holt 1994) carries four
identical grazing-incidence X-Ray Telescopes (XRT, Serlemitsos et
al. 1995; Tsusaka et al.\ 1995).  The point spread function (PSF) of
the XRT has a cusp-shape peak with the half power diameter of
$3'$. Two Solid-state Imaging Spectrometers (SIS0 and SIS1, Burke et
al. 1991) and two GISs (GIS2 and GIS3) are placed at the focal planes
of the four XRTs.  Because of the 3 times larger FOV of the GIS than
the SIS, this paper deals with the GIS data only.  The
GIS covers an energy range from 0.7 to 10 keV, with a spectral
resolution of 8\% FWHM at 6 keV\@.

All the GIS data were obtained in the normal pulse-height (PH) mode.
We screened the data with a telescope elevation angle from the Earth
rim $> 5^{\circ}$, and with an attitude aberration from the average
$< 1'$.  Flare-like events due to background fluctuations (Ishisaki
1996) and data when the spacecraft was in high background regions,
such as the south Atlantic anomaly (SAA), were also excluded.  The
cosmic X-ray background (CXB) was estimated from the archival
blank-sky data which was taken during 1993 $-$ 1994 for a total
exposure time $\sim$ 560 ksec after data screening.  Although the
diffuse soft X-ray emission from the North Polar Spur (NPS), which is
the most prominent part of Loop I (Egger \& Aschenbach 1995), is
located in the south-east direction of the Virgo cluster (Snowden et
al.\ 1995), it was confirmed that its influence on the background
spectrum was negligible in the ASCA energy band (see section 3.1).
The long-term variability of the non X-ray background (NXB) of the GIS
(Ishisaki 1996) was corrected for.  In the spectral analysis, the data
from GIS2 and GIS3 were summed into a single spectrum.

Figure 2 shows the GIS image in the low (0.5 $\sim$ 2.0 keV) and the
high (2.0 $\sim$ 10.0 keV) energy bands.  These images are smoothed by
a Gaussian function with $\sigma = 1.5'$, but no corrections are made
for the background, the vignetting and the point spread function (PSF)
of the XRT.  The X-ray image of almost the whole cluster in the energy
band above 2 keV has been obtained here for the first time. The image
shows a generally similar structure to the low band one, but some hard
sources are clearly seen (e.g.\ NGC4388 and VPC0866).  The X-ray
brightness distribution is strongly peaked at M87 in both energy
bands.  This feature is similar to the luminosity-weighted
distribution of member galaxies, but a simple number-density
distribution of galaxies shows a peak closer to M86 (Binggeli,
Sandage, \& Tammann 1985; Binggeli, Tammann, \& Sandage 1987;
Schindler, Binggeli, \& B\"{o}hringer 1999).  The X-ray emission is
clearly extended over more than $2^{\circ}$ around M87, and several
discrete sources, both background objects and bright member galaxies,
are easily recognized.  In the typical outer region at $2.3^{\circ}$
northwest of M87 (i.e., positions 2002N, 2003N and 2004W), the
intensity of the cluster emission is $0.6 - 0.7$ times that of the total
background level (the sum of the non X-ray and the cosmic X-ray
background).  The linking region between M87 and M49 shows significant
X-ray emission in both energy bands, confirming that the 2 subclusters are
connected with ICM\@.

\subsection{Elimination of Discrete Sources}

Several discrete sources are clearly seen in Figure 2. These sources
have to be removed from the data before we look into the ICM
properties. We prepared a source mask, which masks out the contribution
from contaminating sources, for the spectral analysis.  The 3 bright
galaxies showing extended emission, M87, M86, and M49, are treated
separately by fitting the data with two temperature models.

The source-detection algorithm adjusted for the ASCA mapping data has
been developed for the CXB study by Ueda et al.\ (1999), and applied to
the Virgo data by Kikuchi et al.\ (2000) who detected 231 source
candidates.  However, the strong intensity gradient around M87 and the
systematic contamination due to the stray light of the XRT make direct
application of this method to the present data difficult.  The latest
mapping data obtained in June 1999 (see table 1) and the additional
pointing data near M87 are particularly difficult to analyze with this
method.  On the other hand, the flux level around M87 within $\sim
1.5^\circ$ is about an order of magnitude higher than that in the
south region.  Therefore, the remaining contamination in this region
would be relatively small, even if the discrete-source subtraction is
incomplete.  Because of this, we prepared the source mask by picking up
recognizable sources with the following method for each pointed field.

As a first step, we produced GIS images in three energy bands: 0.5
$\sim$ 2.0 keV, 2.0 $\sim$ 10.0 keV and 0.5 $\sim$ 10.0 keV, for all
60 pointings.  Next, in each image we searched manually for X-ray peaks
and extended excess emission.  For each detected source, the size of
the source mask is determined by the source intensity.  We require
the source mask to satisfy the condition that the remaining flux of the
source outside the mask should be less than $1/3$ of the surrounding
ICM level.  As a result, we produced 50 source masks which could
exclude source candidates with X-ray fluxes $> 5 \times 10^{-13}$ ergs
cm$^{-2}$ s$^{-1}$ in the $2 - 10$ keV band.  All the bright sources
listed in the bright source catalogue of the ROSAT all-sky survey
(Voges et al.\ 1999; Schwope et al.\ 2000) were excluded with this
mask.

The number of sources detected in this work is much smaller compared
with the source candidates reported by Kikuchi et al.\ (2000), who
picked up the sources with a $2 - 10$ keV flux $> 1 \times 10^{-13}$
ergs cm$^{-2}$ s$^{-1}$. To evaluate the goodness of the present
source removal, we estimated the expected fraction of the flux
contamination. As for the GIS, the detection limit with a pointing of
$\sim$ 20 ksec, which is the typical exposure time in the present
mapping observation, is estimated to be $\sim 2 \times 10^{-13}$ ergs
cm$^{-2}$ s$^{-1}$ in $2-10$ keV band (Ueda et al.\ 1999).  Therefore,
we may assume that all the unsubtracted sources have this flux level
in the worst case.  The $\log N - \log S$ relation obtained with the
ASCA (Ueda et al.\ 1999) predicts that the expected number of sources
at this flux level is $\sim 2.5$ FOV$^{-1}$\@. Therefore, the maximum
remaining flux per FOV after the source-mask application becomes $\sim
5 \times 10^{-13}$ erg cm$^{-2}$ s$^{-1}$.  Though the detection limit
increases in the region near M87 where the ICM brightness is much
higher, the expected remaining flux per FOV decreases since the
expected number of sources is small at the high flux level.  The
contaminating fraction, which is the ratio of the remaining flux to
the ICM intensity in the GIS FOV, is less than 5\% in the region
within $3^{\circ}$ from M87 and $\sim 10\%$ in the outer regions. This
flux level is comparable to the field to field fluctuation of the
CXB\@.  The net effect is the reduction of the best-fit ICM
temperatures within the statistical errors, assuming that the
remaining sources have a power-law spectrum with a photon index of
1.7.

To confirm that the resultant ICM temperatures are not significantly
affected by the unsubtracted weak sources, we compared the present
temperature values with the previous results by Kikuchi et al.\ (2000;
northwest and south regions of M87)\@.  The temperature values are
mostly consistent within the statistical error, and there is no
systematic feature such as the correlation with the ICM intensity or
with the distance from M87\@. We conclude that the present
source-detection procedure causes no systematic problems, even in the
faint flux regions.

\subsection{Stray Light Contamination}

Some fraction of the photons from outside of the FOV reaches the focal
plane due to the XRT properties. These photons are called {\it stray
light} and complicate the data analysis for largely extended
sources. It comes from the surrounding sky region within the angular
radius of 2$^{\circ}$ from the center of the FOV\@.  The correct
estimation of the stray light is particularly important in the
analysis of spatially resolved spectral features in the ICM.  We 
evaluate the amount of the stray light by utilizing the ray-tracing
simulation (Tsusaka et al.\ 1995) and include its effect in the XRT
response function used in the spectral analysis. For this estimation,
we first need to know the precise surface brightness distribution of
the cluster.  However, the highly irregular morphology of the Virgo
cluster cannot be approximated by simple analytic models such as the
$\beta$ model.  The ROSAT all-sky survey (RASS; B\"ohringer et al.\
1994) data give a complete image with significantly better angular
resolution than the ASCA\@. Therefore, these data are used here as the
template for the surface brightness profile.  We select the PSPC data
only above 0.5 keV, which gives better matching with the ASCA band. A
spatially uniform background, obtained from a blank sky region, is
subtracted from the Virgo RASS data.

To justify the use of the RASS image, we have to check that the
observed GIS image is well reproduced in the energy range below 2 keV
by the ray-tracing simulation.  For the incident spectrum, we assume a
MEKAL model (Mewe, Lemen, \& Oord 1986) with constant spectral
parameters over the cluster area with a temperature of
2.0 keV and a metal abundance of 0.3 solar, respectively. These values
have been previously derived for the Virgo cluster (see e.g., Koyama,
Takano, \& Tawara 1991; Matsumoto et al.\ 2000).  The actual data
indicate spatial variation in temperature as shown later, and the
ray-tracing code has a systematic error of about 10\%. Allowing for
these errors, we found that the RASS image reproduced the soft-band
GIS image with the satisfactory accuracy. The relative deviation of
the simulated image from the observed one is less than 10\% in all
regions.  We then estimated the stray light effect based on this
template image.

The XRT response calculation based on the RASS image is performed over
a radius of $2^{\circ}$ for each pointed field (coordinates are listed
in table 1).  This size covers the range in which the stray light
causes significant effects.  The XRT response function is computed
with the assumption that the cluster gas is isothermal. This
approximation, which enables us to perform spectral fits for
individual regions, is useful for the first-order estimation of the
spectral parameters (Honda et al.\ 1996).  The ray-tracing simulation
also allows us to trace back the origin of detected photons. We
noticed that the regions $1^\circ \sim 1.2^\circ$ from M87 were
strongly affected by the emission from the cluster center, and the
central flux within $r = 10'$ contributed about 20\% of the detected
flux.  Excluding the regions within about $1^\circ$ from M87, M86 and
M49, almost half of the detected photons are always occupied by the
flux from the pointed sky region in the field of view.

\section{Results}
\subsection{Temperature Distribution}

Pulse-height spectra are produced by accumulating the GIS data in each
pointed field. Most of the regions are tested with single temperature
thermal models (MEKAL) in the energy band $0.8 - 10.0$ keV\@.  The
interstellar absorption is fixed at the Galactic value [$N_{\rm H} =
(1.6 - 2.7) \times 10^{20}$ cm$^{-2}$, Stark et al.\ 1992].  The
plasma temperature, normalization, and metal abundance are the free
parameters.  For three bright galaxies, M87, M86 and M49, we include
an additional soft X-ray component to model the galaxy emission ($kT
\sim 1$ keV; Matsumoto et al.\ 1996; Awaki et al.\ 1994; Matsushita et
al.\ 1994).  The fitted model consists of the 2 components: the hot
component (ICM) described by a 
MEKAL model, and the cool component described either by
the MEKAL or by the MEKAL + Bremsstrahlung model in which the temperature
of the bremsstrahlung is fixed to 10 keV\@. The 10 keV Bremsstrahlung
is known to give an empirical approximation to the spectra for
low-mass X-ray binaries (Makishima et al.\ 1989, Matsushita et al.\
1994, Matsumoto et al.\ 1997).  The two-dimensional distribution of the
ICM temperature derived from the individual spectral fits is shown in
the top-left panel of figure 3.  The top-right and the bottom panels
show projected temperature distribution along the north-south and the
east-west directions, respectively. The errors indicate 90\%
statistical limits.  The best-fit temperatures are summarized in table
1.

As shown in figure 3, the temperature is about 2.5 keV around M87 and
decreases slightly to $\sim 2.0$ keV with a large fluctuation in the
northern region of M87\@.  In the southern region, the average
temperature at $2^\circ$ from M87 is still $\sim 2.5$ keV\@.  In the
linking region between M87 and M49, we confirm the unusually high
temperature recently reported by Kikuchi et al.\ (2000).  The
temperature in the M49 region is $\sim 2.2$ keV and drops to $\sim
1.5$ keV with radius.

The X-ray surface brightness is evaluated for 3 different regions. The
region around M87 with a radius of $0.4^{\circ} - 2.5^{\circ}$
(covered area $\sim$ 12.2 deg$^2$) indicates a 2.0 $-$ 10.0 keV
brightness of ($1.05^{+0.07}_{-0.07}) \times 10^{-14}$ ergs cm$^{-2}$
s$^{-1}$ arcmin$^{-2}$, the M49 region with a radius of 0.4$^{\circ} -
$ 1.0$^{\circ}$ (covered area $\sim$ 2.5 deg$^2$) shows $(1.52
^{+2.05}_{-1.39}) \times 10^{-15}$ ergs cm$^{-2}$ s$^{-1}$
arcmin$^{-2}$, and the region between M87 and M49 with the declination
of $+9^{\circ} 00' \sim +11^{\circ} 00'$ (covered area $\sim$ 2.8
deg$^2$) indicates $(6.33^{+1.79}_{-2.38}) \times 10^{-15}$ ergs
cm$^{-2}$ s$^{-1}$ arcmin$^{-2}$, respectively.

Examples of the spectra are shown in the top panel of figure 4 for the
hot (pos-1508E and 1509E, $\sim 110'$ away from M87) and the cool
(pos-1513N and 2001N, $\sim 120'$ away from M87) regions. The data are
well represented by a single MEKAL model.  The bottom panel shows
the ratios of the observed data and the XRT responses, i.e., the
effective areas of XRT, used in the spectral fit.  Since the responses
are very similar with almost no energy dependence in the energy range
above $\sim 2$ keV, the temperature difference between the two regions
is evident.  The best-fit temperatures are obtained to be
2.65$^{+0.15}_{-0.16}$ keV and 1.71$^{+0.11}_{-0.09}$ keV for the hot
and cool regions, respectively.  The possible origins for the
systematic error of the temperature are the error in the estimated
contaminating spectrum from M87, the remaining contribution from the
masked-out sources, and the fluctuation of the CXB and NXB\@.  The
integrated systematic error on the temperature is estimated to be less
than 0.2 keV, which is much less than the temperature difference
between the 2 regions.

Within $10'$ from M87, a significant variation in the metal abundance
has been reported by Matsumoto et al.\ (1996).  Therefore, the
spectrum in pos-0001, within $\sim 25'$ from M87, cannot be fitted
with a simple two temperature model ($\chi^2_{\nu}$ = 1.71).  Spectral
analysis around the bright M87 region will be presented in a separate
paper (Shibata et al.\ 2000).  

There is a flux contamination from the NPS around M49\@. The PSPC
spectrum was reported by Irwin \& Sarazin (1996) to be $kT = 0.15$ keV
and the $0.5 - 2.0$ keV flux was $ 0.6 \times 10^{-15}$ ergs cm$^{-2}$
s$^{-1}$ arcmin $^{-2}$, respectively. This is only $\sim 4\%$ of the
ICM level in the $0.8 - 2.0$ keV band (used for the spectral analysis)
and $\sim 10\%$ in the $0.7- 2.0$ keV band (for the hardness ratio
analysis in section 3.2), respectively.  Compared with the single
temperature fit, the best-fit ICM temperatures change by about $+0.2$
keV which is within the statistical error in this region.  Considering
also that the affected regions are confined around M49, we may say
that the effect of the NPS is not significant for the present
temperature results.

Figure 5 shows the frequency distribution of the ICM temperature based on
the spectral fits for the GIS FOV unit.  Shaded and unshaded portions
correspond to the data within $2.5^\circ$ centered on M87 and for all
the data in the Virgo field, respectively.  We fitted the distribution
with a Gaussian function and obtained an average temperature of $\bar{T}
= 2.33$ keV with $\sigma_{\bar{T}} = 0.28$ keV within $2.5^\circ$ from
M87 and $\bar{T} = 2.28$ keV with $\sigma_{\bar{T}} = 0.43$ keV for
all regions, respectively. The assumption of a constant temperature
gives a $\chi^2$ value of 351 with 37 degrees of freedom within
$2.5^\circ$ from M87\@.  Therefore, there is a significant temperature
variation over angular scales larger the GIS FOV ($\sim 44'$ or $\sim
260$ kpc).

\subsection{Spatial Variation of the Hardness Ratio}

In this section, we estimate the small-scale ($\sim 5'$) variation of
the ICM temperature based on the hardness ratio analysis.

Hardness ratios (H.R.) are calculated from the background-subtracted
counting rates in the energy bands $2.0 - 7.0$ keV and $0.7 - 2.0$
keV\@. This energy division is chosen to maximize the temperature
sensitivity for the spectra with $kT \sim 2$ keV\@. The top panel of
figure 6 shows the H.R. for each pointed field ($r \sim 22'$) as a
function of the distance from M87 with filled circles.  There is a
small gradient from M87 to M49 and the remarkable hard region between
M87 and M49\@. There is a sharp minimum at $\sim 85'$ from M87, which
is caused by the reflection property of the XRT\@.  This angle
corresponds to the boundary where the single reflection light from the
secondary mirror disappears so that the spectrum becomes softer in
larger angles. The apparent H.R. values are thus strongly dependent on
the XRT response and the surface brightness distribution.
Therefore we need to correct for these effects before converting them
to the temperature.

The conversion from the H.R. values to temperatures is performed
based again on the ray-tracing simulation.  Open circles in the top
panel of figure 6 show simulated H.R. values for an isothermal ICM
with $kT = 2.0$ keV\@.  The assumed spectral model is MEKAL with a
metal abundance of 0.3 solar, and the RASS image is used for the
surface brightness profile.  Since the plasma temperature is a simple
and smooth function of the H.R., the ratio of the observed to
simulated H.R. indicates the correct temperature in each position
(i.e., $kT = 2.0$ keV when the data to model ratio is unity). The
effect of the variation of the metal abundance is such that the
temperature changes by $\pm 0.3$ keV for an abundance change of $\pm
0.2$ solar, respectively.  Since the temperature change is within the
statistical error, the abundance variation does not significantly
affect the temperature results.  The middle panel in figure 6 shows
the corrected H.R. (= H.R.$_{\rm obs.}$/H.R.$_{\rm sim.}$, hereafter
C.H.R.) values which now corresponds to the temperature as indicated
in the right axis.  As a result, the minimum at $\sim 85'$ has
disappeared.  In order to check the adequacy of the above
H.R. temperatures, we again performed the ray-tracing simulation
assuming the obtained distribution of the H.R. temperatures, and
compared the resultant simulated H.R. values with the observed ones, as
shown in the bottom panel.  The dispersion of the C.H.R. has become
considerably smaller than in the isothermal case (middle panel), which
confirms that the the temperature calculation has been consistently
performed.

For the 53 pointed fields represented by the single temperature model,
the correlation between the temperatures derived from the C.H.R. and
from the spectral fit is plotted in Figure 7.  Except for two pointed
fields (pos-5002S and 5003S, open square) which indicate very low
metal abundance ($<$ 0.1 solar) and low temperature ($<$ 1.3 keV)
compared with the model spectrum for the simulation, the temperature
values agree well.  If we fit the correlation with a linear function,
a slope of $0.99 \pm 0.01$ is obtained with a $\chi^2_{\nu}$ value of
0.72 (d.o.f.\ = 51).  Based on this agreement, we look into the
small-scale temperature variation using the C.H.R.

The two-dimensional map of the temperature is calculated from the
C.H.R. and shown in Figure 8. The C.H.R. is calculated for a $20'
\times 20'$ square region, and the center of its integrated region is
shifted by a $5'$ step in order to raise the photon statistics.
Therefore, the temperature map essentially indicates running means for
the temperature variation.  The XRT response and the vignetting effect
are taken into account.  The source mask mentioned in section 2.2 is
also applied.  Since three bright galaxies M87, M86 and M49 contain
a cool spectral component, these regions indicate lower H.R. values.
The one-dimensional variation of the C.H.R. is shown in figure 9 along
the 5 paths indicated in the right panel. This plot also shows the
errors in temperature. The 5 paths shown here are as follows; (A)
connects the positions of M87 and M49, (B) runs through the high
temperature region in the linking region of M87 and M49, (C) is the
west-east cut passing through M87, (D) includes the high temperature
region at 1.7$^{\circ}$ north-west of M87, and (E) connects M87 and
M86, respectively.

Figure 8 and 9 show that the region within about $60'$ from M87 has
little variation of the ICM temperature, except for the inner region
($r \sim 15'$) in which the cool ISM component exists.  On the other
hand, the outer region shows a significant variation of temperature.
As already seen in the H.R. feature in figure 3, we can recognize that
the periphery of M87 is hotter than the M49 region, and the linking
region between both galaxies shows a significantly harder emission.

\subsection{Spatial Scale of Temperature Variation}

As shown in figure 8 and 9, we have
detected significant temperature variation in the ICM\@. The data
enable us a quantitative evaluation of the spatial scale for the
temperature variation. We will investigate it by calculating the
auto-correlation function (ACF)\@. The cell size is $10' \times 10'$,
and the temperature is calculated from the C.H.R. as described in the
previous section. The ACF is calculated for the two dimensional
distribution of the C.H.R., as a function of the
angular distance $\theta$ between two cells. This is given by the
following formula:
\[
W(\theta) ~=~ 
\left <
~\frac{\left \{ S({\it \bf r}_{i}) - \bar{S} \right \}
\left \{ S({\it \bf r}_{j}) - \bar{S} \right \} }{\bar{S}^2}
\right >_{\theta_{ij} = \theta} ~~~~
(i \leq j, ~~i = 1,2, \ldots, N)
\]
\[
\left ( ~~
\theta _{ij} ~=~ \left |  {\it \bf r}_{i} - {\it \bf r}_{j}
\right | 
~~ \right )
\]
where $S({\it \bf r}_{i})$ is the C.H.R. at the
$i$-th position, $\bar{S}$ is 
the average ratio ($= 1.20$
corresponding to 2.43 keV),
$N$ is the total cell number, 
and $<~>$ denotes an angular average, 
respectively.
If we assume that the hardness ratio is a random variable obeying a
Gaussian distribution with a mean of $\bar{S}$ and variance of
$\sigma_{S}^{2}$, the standard deviation of the ACF is given by
\[
\sigma_{W(\theta)} = 
\frac{1}{\sqrt{N}} \left ( \frac{\sigma_{S}}{\bar{S}} \right )^2 ~~.
\]

In this calculation, we exclude the regions showing the two
temperature components: near M87, M86 and M49\@. The resultant ACF is
plotted in figure 10 (filled circle).  A significant correlation
exists within an angular distance of $\sim 1.5^{\circ}$. On larger
angular scales, the correlation strength drops to almost zero and
shows no features.

There could be a systematic effect due to the large-scale temperature
gradient centered on M87, as recognized in figure 3.  This gradient
can be represented by a power-law function as $T(r)$ = 3.83 $\times ~
r^{-0.12}$ [keV] for a radius range $30' \sim 150'$.  The expected ACF
profile with this gradient is shown with the broken line in figure 10.
The feature due to the temperature gradient is too small to
account for the ACF profile presently obtained for the Virgo cluster,
in particular on angular scales of $\sim 1^{\circ}$ ($\sim$ 350
kpc).

The obtained auto-correlation feature can be interpreted in the
following way.  The strong correlation in $\theta < 1^{\circ}$
suggests that up to this scale the ICM tends to have the same
temperature.  This angular scale corresponds to an actual size of 300
kpc at the Virgo cluster. On larger angular scales, the temperature
distribution seems to be random within the statistical error. We note
that significant temperature fluctuations with a scale of individual
GIS fields ($\sim 44'$ or $\sim 260$ kpc) as seen in figure 5
 is consistent with the auto-correlation
result.

\subsection{Correlation with Galaxy Properties}

The ICM temperatures are compared here with the properties of galaxies
available in the VCC (Binggeli, Sandage, \& Tammann 1985, Binggeli \&
Cameron 1993), which gives us number counts, average heliocentric
velocities, and velocity dispersions ($\sigma_r$) of the local member
galaxies.  We adopt the ICM temperature evaluated in each pointed
field (shown in section 3.1).  The galaxy properties are mean values
in a radius of $1.5^\circ$ centered on the GIS-FOV, thus averaging is
here over a larger field than for the temperature, which is necessary
to obtain enough statistics.

Galaxy counts and average velocities both indicate a weak positive
correlation with the ICM temperature.  We recognize that cooler
regions tend to show a lower galaxy count and a lower velocity than
other regions.  No significant correlation is seen for the
luminosity-weighted galaxy count, however.

The left panel of figure 11 shows a correlation between the
temperature and the velocity dispersion in the whole cluster. The
symbols indicate different regions. Triangles are for the northwest of
M87 (around M86), filled circles are within $2.5^\circ$ from M87
except for the northwest region, and open circles are the remaining
regions, respectively.  The clustering of the same symbols suggests
that galaxies form distinct groups. In particular, the group around
the velocity dispersion of $\sim 800$ km s$^{-1}$ is clearly different
from others.

The presence of a distinct group is also suggested by the $\beta_{\rm
spec} (= \mu m_p \sigma_r^{~2} / kT$) value drawn in the left panel of
figure 11. This is the ratio between the velocity dispersion of
galaxies and the thermal energy of the ICM.  The general cluster
region indicates a typical value of $\beta_{\rm spec} \sim 1.0\pm0.5$,
and the northwest region shows unusually large $\beta_{\rm spec}$
values above 2.0.  If galaxies and gas are in dynamical equilibrium,
we expect $\beta_{\rm spec}$ to be about unity.  The map of
$\beta_{\rm spec}$ values in the right panel of figure 11 shows that
these high $\beta_{\rm spec}$ values are distributed around M86 and in
the northern direction.  This indicates that the ICM has not reached
dynamical equilibrium, which is probably related to a subcluster
merger in the M86 region.

\subsection{Inspection of the Extended Non-thermal Emission}

Recently, the advanced sensitivity of BeppoSAX for extended hard X-ray
emission above 10 keV has enabled the detection of non-thermal emission
from several clusters (Coma cluster; Fusco-Femiano et al.\ 1999,
A2199; Kaastra et al.\ 1999, A2256; Fusco-Femiano et al.\ 2000).
Since the Virgo cluster has a relatively low ICM temperature ($\sim 2.5$
keV), the thermal emission rapidly decreases above 6 keV, as shown in
the bottom panel of figure 12.  This allows us to search for extended
non-thermal emission with the ASCA GIS in the energy band below 10
keV\@.

Spectral data were summed up, and the overall spectrum of the ICM was
produced by excluding several exceptional regions.  The pointing on
M87 (pos-0001, $r < 22'$) shows extremely strong emission
characterized by two temperature component spectra and has been
removed from the average spectrum. The hot region between M87 and M49
and the region around M86 and M49 show significant deviation from the
average temperature of about 2.5 keV, and have been excluded.  As a
result, the average Virgo spectrum is produced for a radius of
$0.4^{\circ} - 2.5^{\circ}$ from M87 with a covered area of $\sim
12.2$ deg$^2$. This includes 36 pointing data with a temperature of
$2.2 \pm 0.5$ keV after the source mask was applied as mentioned in
section 2.2.

The combined spectrum was first fitted with the MEKAL model in the
energy range $0.8 -10.0$ keV\@. Concerning the very high statistics,
systematic errors need to be included in the fit.  The current
response of the GIS requires a systematic error of 0.5\% to obtain
an acceptable fit for the spectral data of the Crab nebula. Therefore,
we include the same systematic error in fitting the overall Virgo
spectrum.  The plasma temperature, normalization, and metal abundances
of Fe, S and Si are free parameters. The interstellar absorption
is fixed at the average Galactic value in this direction ($N_H = 2.55
\times 10^{20}$ cm$^{-2}$).  A single temperature model gives a 
best-fit $\chi^2_{\nu}$ value of 1.23 (d.o.f.\ = 322), and a 
two-temperature model yields $\chi^2_{\nu} =1.10$ (d.o.f.\ = 320). The
two-temperature model is acceptable at the 90\% confidence limit.  The
best-fit temperatures are $3.67 ^{+0.53}_{-0.52}$ keV and $1.38
^{+0.09}_{-0.13}$ keV, respectively, and the metal abundances are
$Z_{\rm Fe} = 0.16 ^{+0.04}_{-0.03}$ solar, $Z_{\rm S} = 0.52
^{+0.10}_{-0.09}$ solar, and $Z_{\rm Si} = 0.33 ^{+0.08}_{-0.06}$
solar, respectively. This two-temperature model should be regarded as
a simple empirical representation of the ICM component to put
constraints on the non-thermal component.  In other words, it is the
most simple model which can represent multi-temperature components
shown in figure 3 and 8 within the photon statistics.  Therefore, only
the Fe abundance which strongly depends on the intensity of L-$\alpha$
emission line is somewhat uncertain.  The parameter values are
summarized in table 2, and the fitted spectrum is also shown in the top
panel of figure 12.  The X-ray flux in the integrated region, already
mentioned in section 3.1, is $(1.09 ^{+0.23}_{-0.14}) \times 10^{-14}$
ergs cm$^{-2}$ s$^{-1}$ arcmin$^{-2}$ in the $2.0 - 10.0$ keV band.

Next, we include in the fit a power-law component to examine the
possibility of non-thermal emission. The model consists of the
two-temperature MEKAL and the power-law (PL) components.  Since a 
fit with a free power-law index $\Gamma$ could not constrain the
parameter, we assume it to be 1.7.  The fitting results are
summarized in table 2. The $\chi^2$ value shows no improvement by the
addition of the PL component.  The upper limit of the PL flux is $
9.52 \times 10^{-16}$ ergs cm$^{-2}$ s$^{-1}$ arcmin$^{-2}$ in the
$2.0 - 10.0$ keV band, which corresponds to less than 9\% of the total
flux.  Figure 12 shows the spectral components (thermal component and
the upper limit of the PL component) together with the CXB and the NXB
spectra.  The BGD components become dominant above 6 keV\@.  We
evaluate the influence of the background uncertainty to the upper
limit of the PL component.  Since the uncertainty in the background
spectra are estimated to be $\sim 10\%$ (CXB) and 6\% (NXB) for the
single GIS FOV, the uncertainty for the combined Virgo spectrum ($\sim
31$ GIS FOV) is estimated to be 1.8\% for CXB and 1.1\% for NXB,
respectively.  If we take the lowest possible NXB flux, the upper
limit on the PL component becomes 14\% larger than the previous result
($1.09 \times 10^{-15}$ ergs cm$^{-2}$ s$^{-1}$ arcmin$^{-2}$).

In figure 12, we recognize that the upper limit of the PL component is
lower than the CXB level by a factor of $\sim 8$.  The upper limit for
the PL flux and luminosity for the whole integrated region are $4.78
\times 10^{-11}$ ergs cm$^{-2}$ s$^{-1}$ and $2.28 \times 10^{42}$
ergs s$^{-1}$ in 2 $-$ 10 keV, respectively. If we attribute this flux
to contaminating sources fainter than the detection limit ($\sim 2
\times 10^{-13}$ ergs cm$^{-2}$ s$^{-1}$), the number of the sources
should be about 240.  However, the $\log N - \log S$ relation (Ueda et
al.\ 1999) indicates that only 30\% of the PL flux can be due to
unresolved contaminating sources.  On the other hand, the integrated flux
from low mass X-ray binaries (LMXB) in the member galaxies can
contribute to the hard X-ray spectrum. The optical Luminosity ($L_B$)
in the integrated region ($0.4^\circ - 2^\circ.5$ from M87 excluding
M86) is estimated to be $1.07 \times 10^{12}$ $L_{\odot}$ from the VCC
data (Binggeli, Sandage, \& Tammann 1985). Because the discrete-source
flux is proportional to the stellar content given by the relation of
$L_X^{\rm LMXB}$ / $L_B \sim 10^{-3.81}$ (Matsushita, 2000), the total
flux from LMXBs is $6.38 \times 10^{41}$ ergs s$^{-1}$, which
corresponds to about 28\% of the upper-limit PL flux.

By subtracting these known contributions, the remaining PL component
can have a luminosity of $9.12 \times 10^{41}$ ergs s$^{-1}$ in the
2 $-$ 10 keV band. This is the upper limit for the diffuse emission in
the Virgo cluster.  Compared with the recently reported hard X-ray
emission from some clusters (e.g., Coma cluster; Fusco-Femiano et al.\
1999, A2199; Kaastra et al.\ 1999, A2256; Fusco-Femiano et al.\ 2000),
the present upper limit for the Virgo cluster is substantially lower
by $1- 2$ orders of magnitude.

\section{Discussion}
\subsection{Temperature Variation and Heating Mechanism of ICM}

Significant temperature variation has been observed in the Virgo
cluster, as shown in section 3.1 and 3.2.  It is characterized by a
systematic increase of the fluctuation range at larger radii (figure
3).  The amplitude of the temperature variation when the data are
integrated in the GIS FOV (a diameter of $\sim 44'$ or $\sim 260$ kpc)
is $> 2$ keV, which is significantly larger than the statistical
fluctuation (figure 5).  The temperature variations in finer scales
are also examined based on the H.R. as shown in figure 8 and 9.  In
figure 10 the typical scale of the H.R. is estimated to be $<
1^{\circ}$ based on the auto-correlation analysis.  These features
indicate that the variation occurs with a crude spatial scale of $\sim
300$ kpc.  In addition to the irregular morphology in the optical and
X-ray band, the temperature distribution has shown a new aspect of the
Virgo cluster in that the ICM is undergoing a strong dynamical
evolution.

Let us look into the heating and cooling mechanisms of the ICM based
on the temperature structure.  We assume that the heating is caused by
an infall of gas into the gravitational potential or by subcluster
mergers.  It is believed that the formation and evolution of rich
clusters takes several $10^9$ yr which is a significant fraction
of the Hubble time. The time scales relevant for the gas heating are
relaxation times for collisions for proton-proton (p-p),
proton-electron (p-e) and electron-electron (e-e), respectively, in
which the p-e collision takes the longest relaxation time
(see Sarazin 1988). This is
given by,
$t_{\rm eq(p,e)} \approx 1 \times 10^8 ~~{\rm yr}~~
(T_{\rm gas} / 2.5 ~{\rm keV}) ^{3/2}
( n / 1 \times 10^{-3} ~{\rm cm}^{-3})^{-1}$
, where $T_{\rm gas}$ is the average ICM temperature ($\sim$ electron
temperature) and $n$ is the typical ICM number density around M87\@.
Therefore, the ICM temperature is raised in $\sim 10^8$ yr after
the initial heating. The cooling time is given by,
$t_{\rm cool} \approx 5 \times 10^{10} ~~{\rm yr}~~
( n / 1 \times 10^{-3} ~{\rm cm}^{-3}) ^{-1}
( {T_{\rm gas} / 2.5 ~{\rm keV}}) ^{1/2}$, 
indicating that the cooling is negligible except for the central
region of the cluster. Therefore, X-ray temperature (electron
temperature) is a good indicator for the local plasma temperature.

The X-ray surface brightness mainly indicates the plasma density,
since the variation of emissivity is a slow function of the 
temperature at $kT
\sim 2$ keV\@. The time scale for a local density fluctuation to be
smoothed out is determined by the sound crossing time,
\[
t_{\rm sound} \approx 4 \times 10^8 
\left ( \frac{T_{\rm gas}}{2.5 ~{\rm keV}} \right ) ^{-\frac{1}{2}}
\left ( \frac{D}{300 ~{\rm kpc}} \right ) 
 \; {\rm yr},
\]
where $D$ is the linear distance of the temperature variation. This
indicates that a gas density (or surface brightness) structure
disappears in much faster than $1 \times 10^9$ yr. The temperature
variation, on the other hand, would be smoothed out with the time scale
of thermal conduction. This is given by,
\[
t_{\rm cond} 
  \approx  
 2 \times 10^9 
 \left( \frac{n}{1 \times 10^{-3}\; {\rm cm^{-3}}} \right)
 \left( \frac{T}{2.5\; {\rm keV}} \right)^{-\frac{5}{2}}
 \left( \frac{{\rm ln\;} \Lambda}{40} \right)
 \left( \frac{D}{300\; {\rm kpc}} \right)^2
 \; {\rm yr},
\]
where $\ln \Lambda$ is the Coulomb logarithm which is nearly
independent of density or temperature for $T_{\rm e} > 4 \times 10^5$
K\@. This is much longer than the sound crossing time, and it is
reasonable that temperature variation persists even though no
structure is seen in X-ray morphology.

The observed image of the Virgo cluster is very smooth.  B\"ohringer
et al.\ (1994) could fit the ROSAT radial profile with a single
$\beta$ model with an exception only in the southern region. Since the
volume emissivity is proportional to $n^2\sqrt{T}$, the plasma
density should be very smooth in the whole cluster. This means that in
most of the region the density disturbance should have occured more
than $4 \times 10^8$ yr ago.  On the other hand, the presence of the
high and low temperature regions indicate that these regions have been
heated up less than about $2 \times 10^9$ yr ago. This time scale is
rather short compared with the Hubble time, and we may conclude that the
Virgo cluster is still evolving with local gas heating occurring in
the cluster outskirts which have low ICM density.

The estimated distance scale for the temperature variation ($\sim 300$
kpc) is larger than the size of typical galaxies, and close to that of
galaxy groups.  It seems reasonable to suppose that the gas and
galaxies have formed certain clumps of this size before falling into
the gravitational potential of the main cluster. Numerical simulations
of the bottom-up structure formation predict just the same features
(Frenk et al.\ 1996).  The high temperature regions probably trace the
heating front where small groups have recently encountered and merged
with the ICM in the main cluster.

The Virgo cluster provides us with the best close-up view of 
ICM features on 100 kpc scales.  It would be difficult
to resolve such features in more distant clusters because of the
angular resolution and limited statistics available with
ASCA\@. However, ASCA and ROSAT observations have detected spatial
temperature structures from many clusters over several hundred kpc to
1 Mpc scales.  Significant asymmetric temperature structures are
observed from the following systems: A115 (Shibata et al.\ 1999), A754
(Henrikson \& Markevitch 1996), A1367 (Donnelly et al.\ 1998), Coma
cluster (Honda, et al.\ 1996, Watanabe et al.\ 1999), Centaurus cluster
(Churazov et al.\ 1999), A2142 (Henry \& Briel 1996), A2255 (Davis \&
White 1998), A2319 (Molendi et al.\ 1999), Cygnus A cluster, A3667,
A2065 (Markevitch et al.\ 1999), A85 and A2657 (Markevitch et
al. 1998). Note that these results mostly correspond to large-scale
mergers in clusters.  Therefore, small-scale (a few hundred kpc)
temperature variations such as seen in the Virgo cluster may well be
missed in the past observations. It is likely that such small
temperature structures would exist in many clusters, in particular in
the outer regions where the conduction time scale is long. These
features, if observed in future, would provides us with important
knowledge about the evolution process of clusters.

\subsection{M86 and M49 Subclusters}

The $\beta_{\rm spec}$ distribution shown in figure 11 indicates that
the ICM and member galaxies have not reached dynamical equilibrium in
some regions.  In the previous section, we discussed that the gas
heatings could have occurred through infalls of group-size blobs.
Here, we examine X-ray and optical properties of 2 known subclusters
in the Virgo system, M86 and M49, in some detail.

M86 is located in the northwest of M87 with an angular separation of
$\sim 80'$ ($\sim 0.3$ Mpc in projection). However, it is more distant
than M87 by $1.2 \pm 0.6$ Mpc based on the measurement of the
planetary nebula luminosity function (Jacoby et al.\ 1990). The Virgo
system itself is considered to have an enormously large depth in the
line of sight ($15 \sim 20$ Mpc, Fukugita, Okamura, \& Yasuda 1993,
Yasuda, Fukugita, \& Okamura 1997). The relative velocity between the
single galaxies, M87 and M86, is estimated to be $\sim 1500$ km
s$^{-1}$ (Binggeli, Sandage, \& Tammann 1985). When the heliocentric
velocities are averaged over the M87 and M86 subclusters, within $60'$
from each galaxy, the 2 subclusters indicate $\sim 1100$ km s$^{-1}$
for M87 and $\sim 600$ km s$^{-1}$ for M86, respectively.  Because the
subcluster average inevitably includes some overlapping galaxies in
the boundary, we should take 500 km s$^{-1}$ as the minimum velocity
for the approaching speed of M86 to M87\@. We will, therefore, use a 
velocity of 1500 km s$^{-1}$ for the later estimation.

If these 2 subclusters are still separated in the line of sight, we
expect the velocity distribution in the direction of the M86
subcluster to show 2 peaks because the sample should contain
foreground M87 galaxies.  However, the observed distribution shows a
single broad profile with a flat top.  This suggests that some part of
the M86 subcluster has already merged into the M87 subcluster.  In
this case, the resultant velocity dispersion becomes very large
because the subcluster bulk motion is added to the original
dispersion. This seems to be the case around M86, which shows a large
velocity dispersion compared with other Virgo fields (left panel of
figure 11).  Also, the ROSAT observation of M86 shows a sharp drop in
X-ray the surface brightness on the M87 side (Rangarajan et al.\
1995).  This apparent compression of the M86 interstellar matter may
be caused by the interaction with the M87 subcluster.  These features
strongly suggest that M87 and M86 subclusters are already interacting
with each other.

The ICM temperature is expected to rise due to the subcluster merger.
For simplicity, we assume that the total kinetic energy of the
subcluster is fully converted to the thermal energy. Using the
relative colliding velocity $v_{\rm col}$, the resultant temperature
increase is given by,
\[
k \Delta T = \frac{1}{3} \mu m_{\rm p} v_{\rm col}^2
           \simeq 4.7 ~~{\rm keV} ~\left ( \frac{v_{\rm col}}
{1500 {\rm km ~s^{-1}}} \right )^2 
\]
where $\mu$ is the mean molecular weight (0.6) in amu, and $m_{\rm p}$
is the proton mass, respectively.  The ASCA results do not indicate
such a high temperature around M86\@.  Even if we may assume that the
high temperature component, once created, is screened out by the large
depth of the Virgo ICM, we can not explain the absence of excess X-ray
brightness especially in the region between M87 and M86\@.  Therefore
it seems more reasonable to suppose that the M86 subcluster is in the
very early stage of a merger and that the heating of the ICM is just
about to occur.

Contrary to the large velocity dispersion, the ICM temperature in the
M86 region is similar to or lower than the other regions at the same
distance from M87\@. This pushes $\beta_{\rm spec}$ unusually high
above 2.0 (right panel of figure 11). In the early stage of a
subcluster merger the velocity distribution naturally creates a broad
profile due to the bulk motion, even before the actual collision takes
place. In fact, N-body and hydrodynamic numerical simulations indicate
that the velocity dispersion increases slowly at first before the rise
of the ICM temperature (Takizawa 1999).  Therefore, the observed high
$\beta_{\rm spec}$ value around M86 is likely to indicate the
beginning of the merger.  Similar high $\beta_{\rm spec}$ values above
2.0 are seen in other northwest regions, suggesting that the infall of
the gas into the gravitational potential has not fully generated the
thermal energy in this region.  The lower average temperature and the
large fluctuation suggest that the dynamical process is going on in
this part of the Virgo cluster, most clearly represented by the M86
subcluster.

Another large subcluster, M49 system, has a similar geometrical
configuration to the M86 one.  Its location is about 1.2 Mpc south of
M87 and further away by $0.6 \pm 0.7$ Mpc.  From the ROSAT
observation, the X-ray contours of M49 are elongated in the
northeast-southwest direction, and a bow shock-like structure is
evident in $\sim 4'$ north of the M49 center.  The relative velocity
at this position, estimated from a pressure balance between the ISM of
M49 and the ICM of the Virgo cluster, is $\sim 1300$ km s$^{-1}$ which
is a reasonable value for the infall velocity of M49 into the cluster
center (Irwin \& Sarazin 1996).  ASCA observations have revealed a
remarkable hot region with $kT \sim 4$ keV in the middle of M87 and
M49 (Kikuchi et al.\ 2000), refining the previous Ginga results
(Koyama, Takano, \& Tawara 1991). In the case of M49 subcluster, the
shock heating seems to have already started. However, it is puzzling
that we do not find a high temperature component in the more closer
region to M49, where the shock feature is suggested.  One possibility
is that our observing angle is rather close to the face-on
configuration to the shock front. In this case, the temperature
averaged over the line-of-sight does not show a clear edge, which is
also demonstrated by numerical simulations (Takizawa 2000). Therefore,
it is likely that the M49 subcluster is in a somewhat advanced stage
of a merger process with M87, compared with the M86 case.

\subsection{Non-thermal Emission}

Based on the best statistical data so far available for the Virgo
cluster, we have searched for the non-thermal emission in the combined
GIS spectrum and detected no significant hard X-ray component with an
upper limit of $L_X^{\rm hard} \sim 9 \times 10^{41}$ ergs s$^{-1}$ in
the $2 - 10$ keV band. This is comparable to the contribution from
contaminating sources or from discrete X-ray sources in the member
galaxies within a factor of 1.5. The expected amplitude of the CXB
fluctuation is $\sim 2 \times 10^{41}$ ergs s$^{-1}$ for the average
of the integrated fields.  The reported luminosities of the
non-thermal emission from Coma cluster, A2199, and A2256 are $0.3 -
1.1 \times 10^{44}$ ergs s$^{-1}$ in the energy band $2- 10$
keV\@. Therefore, the upper limit for the Virgo cluster is lower by $1
-2$ orders of magnitude, suggesting that the non-thermal activity over
the cluster scale is almost absent in this cluster. The upper limit on
the luminosity is comparable to the hard X-ray emission from the
galaxy group HCG62, which clearly shows an excess hard X-ray component
in the ASCA data ($\sim 8 \times 10^{41}$ ergs s$^{-1}$; Fukazawa et
al.\ 2000). Considering that all the hot regions recognized in the
Virgo cluster indicate the spatial size of galaxy groups, it may be
that the non-thermal activities occurring in this cluster are confined
to regions of the size of groups. In this sense, the Virgo cluster is
a young unevolved system, and more large-scale gas heatings would
probably take place in future.

\section{Conclusion}

The Virgo cluster is the nearest system and enables us to study of
the temperature distribution with sufficient spatial resolution by
ASCA\@. We have obtained the full temperature map of this cluster for
the first time, and found significant variation of the temperature with a
spatial scale of about 300 kpc. This feature is probably an evidence
that heating of the ICM proceeds with infall or mergers of gas
clouds which have the size of galaxy groups, as expected from the
current scenario of the structure formation. Comparison of the thermal
energy of the ICM and the velocity dispersion of member galaxies
 have been carried out for the whole Virgo system. We have reached a
picture that the M86 and M49 subclusters are in a different stage of
subcluster mergers, with M86 just in the pre-heating phase and M49
already entered a shock heating phase. We have not detected extended
non-thermal emission from the Virgo cluster.  Even the upper limit is
much lower than those of other rich clusters (Coma cluster, A2199, and
A2256), but is comparable to that of the galaxy group HCG62.  We
expect that future X-ray observations would detect more evidences of
such a local heating of the ICM in other clusters and give us a
comprehensive view of the actual process of the cluster evolution.

\acknowledgments

We would like to express our thanks to the ASCA team and the
ASCA\_ANL, SimASCA, and SimARF software development team for
constructing excellent frameworks for the analysis.  We also thank Dr. 
Y. Ikebe, T. Reiprich, M. Takizawa and H. Inoue for valuable advice on
the analysis method and stimulating discussions.  Useful comments from
the referee Dr.\ J. Irwin are gratefully acknowledged.
R. S. acknowledges support from the Japan Society for the Promotion of
Science (JSPS) for Young Scientists.  This work was supported in part
by a Grant-in-Aid for Scientific Research No. 12304009 from JSPS.

\clearpage

\clearpage
\begin{figure}[h]
\begin{center}
\mbox{\psfig{figure=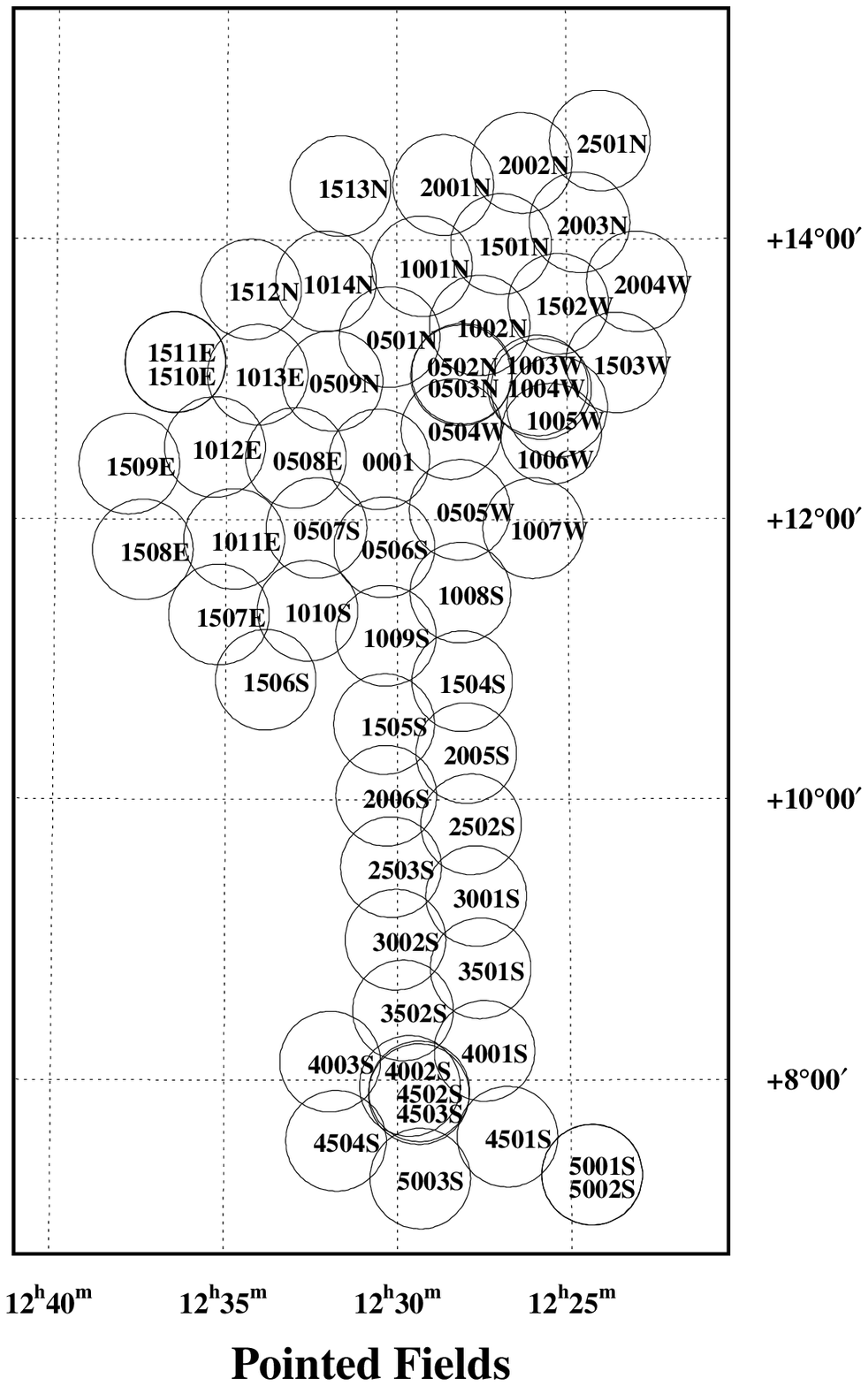,width=0.73\textwidth,angle=0}}
\caption{ 
Pointed fields observed with ASCA. The observational parameters are
listed in table 1. Radius of each circle is 22$'$
corresponding to the GIS field of view.
}
\end{center}
\end{figure}

\begin{figure}[h]
\begin{center}
\mbox{\psfig{figure=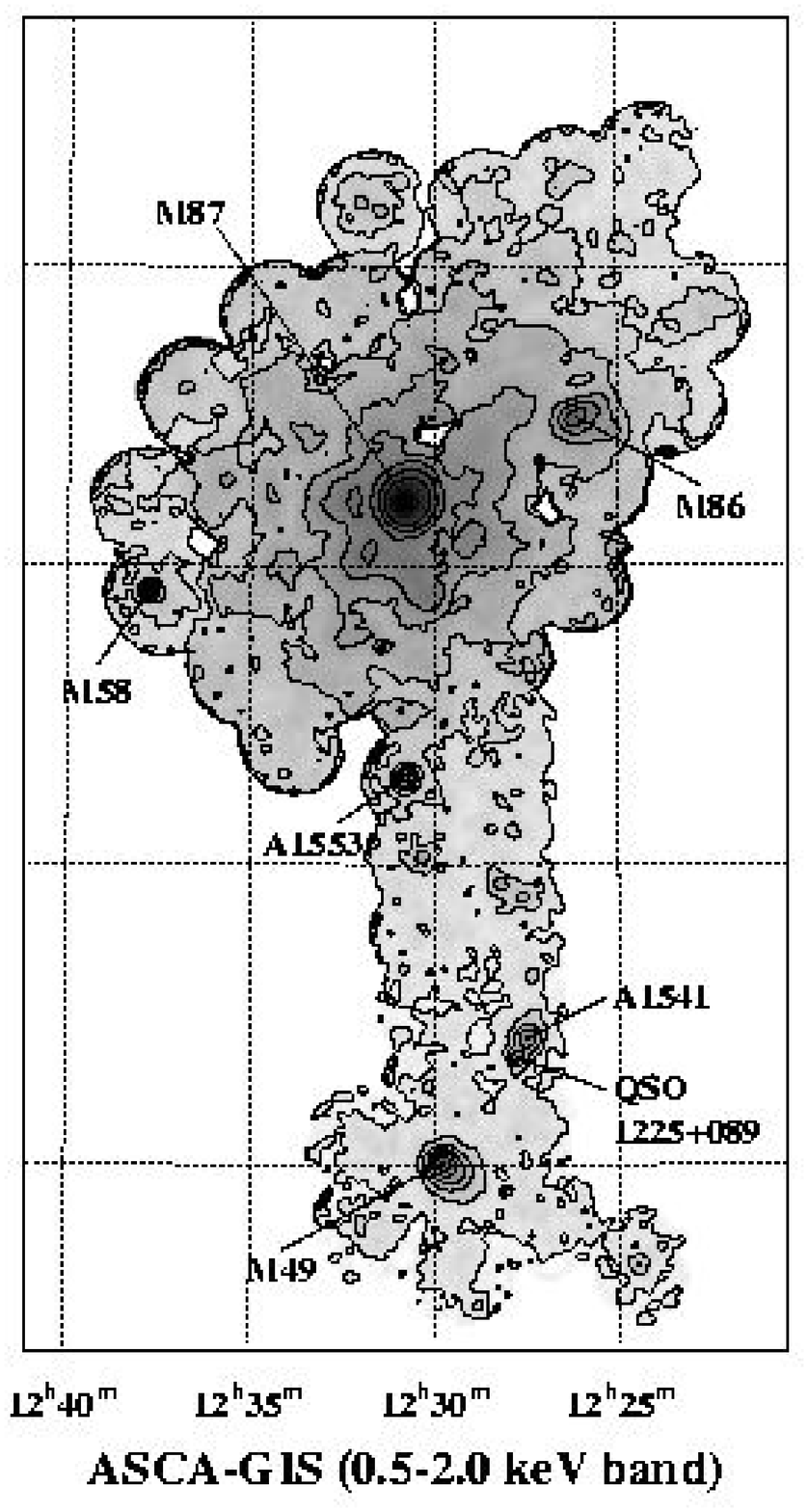,height=0.65\textheight,angle=0}}
\mbox{\psfig{figure=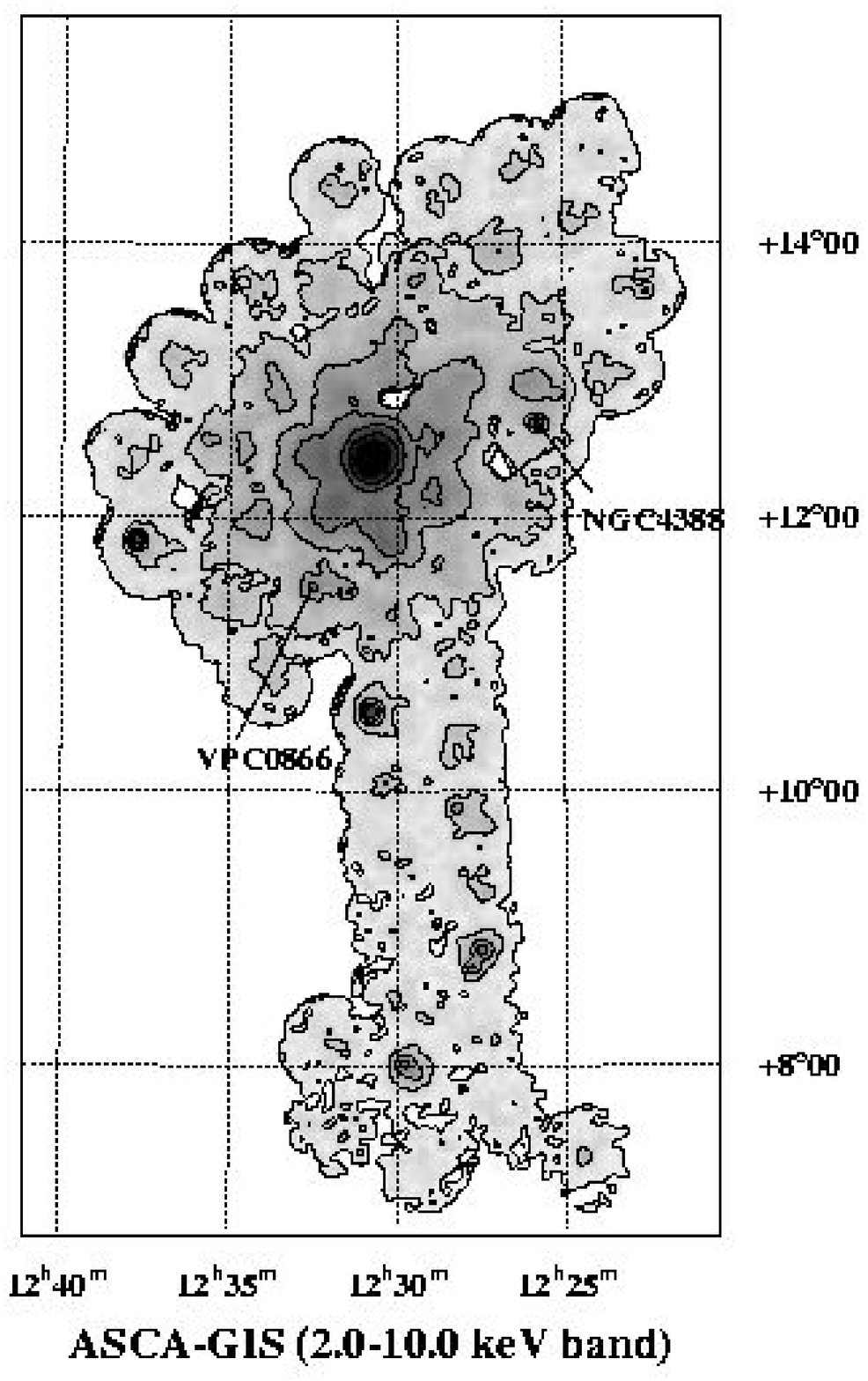,height=0.65\textheight,angle=0}}
\caption{
({\it left}): A 0.5 $-$ 2.0 keV band image of the Virgo cluster.
({\it right}): A 2.0 $-$ 10.0 keV band image of the Virgo cluster.
Each image is smoothed by a Gaussian function with $\sigma$ = 1.5$'$,
with background and vignetting effect not corrected for. Contour
levels are logarithmic scale with a step factor of 1.78.
}
\end{center}
\end{figure}

\begin{figure}[h]
\begin{center}
\vspace*{-6mm}
\mbox{\psfig{figure=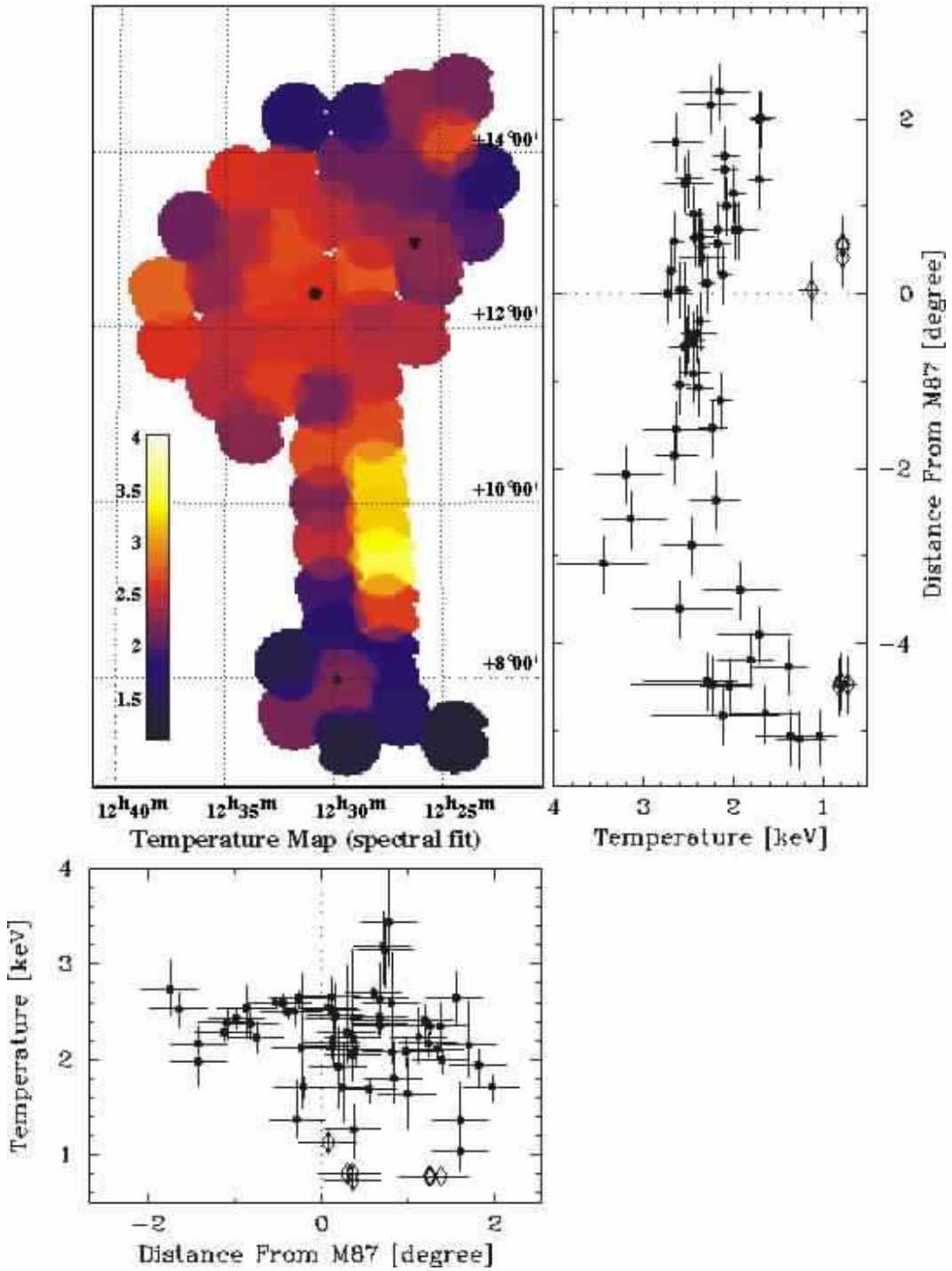,width=0.85\textwidth,angle=0}}
\caption{
ICM temperature map of the Virgo cluster, estimated by the spectral
fits for each GIS field of view.  ({\it top-left}): Two dimensional
temperature map.  The positions of three galaxies are marked: M87
(circle), M86 (inverse triangle), M49 (triangle).  ({\it top-right}):
Projected temperature distribution in the north to south direction.
({\it bottom}): Projection in the east to west direction.  Diamonds
indicate the cool component (i.e., ISM) in M87, M86 and M49.
}
\end{center}
\end{figure}

\begin{figure}[h]
\begin{center}
\mbox{\psfig{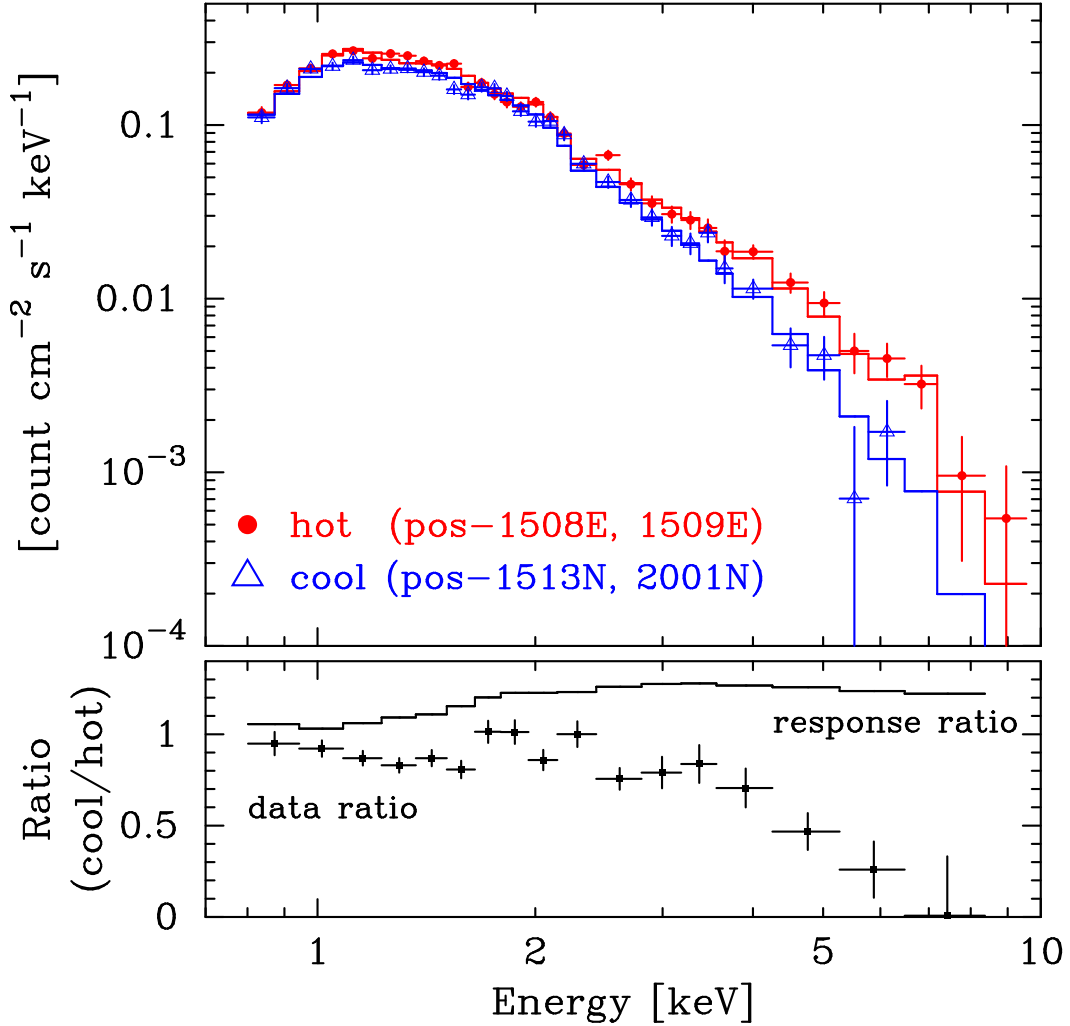}}
\caption{
({\it top}): Observed GIS spectra of the hot (pos-1508E and 1509E,
filled circle) and cool (pos-1513N and 2001N, open triangle) regions. 
The best-fit spectra of the MEKAL model are shown as solid lines.
Best-fit temperatures are 2.65$^{+0.15}_{-0.16}$ keV and
1.71$^{+0.11}_{-0.09}$ keV for the hot and cool regions, respectively.
({\it bottom}): Ratios of the observed data and the XRT response
functions (i.e., effective areas of XRT) for spectral analysis.
}
\end{center}
\end{figure}

\begin{figure}[h]
\begin{center}
\mbox{\psfig{figure=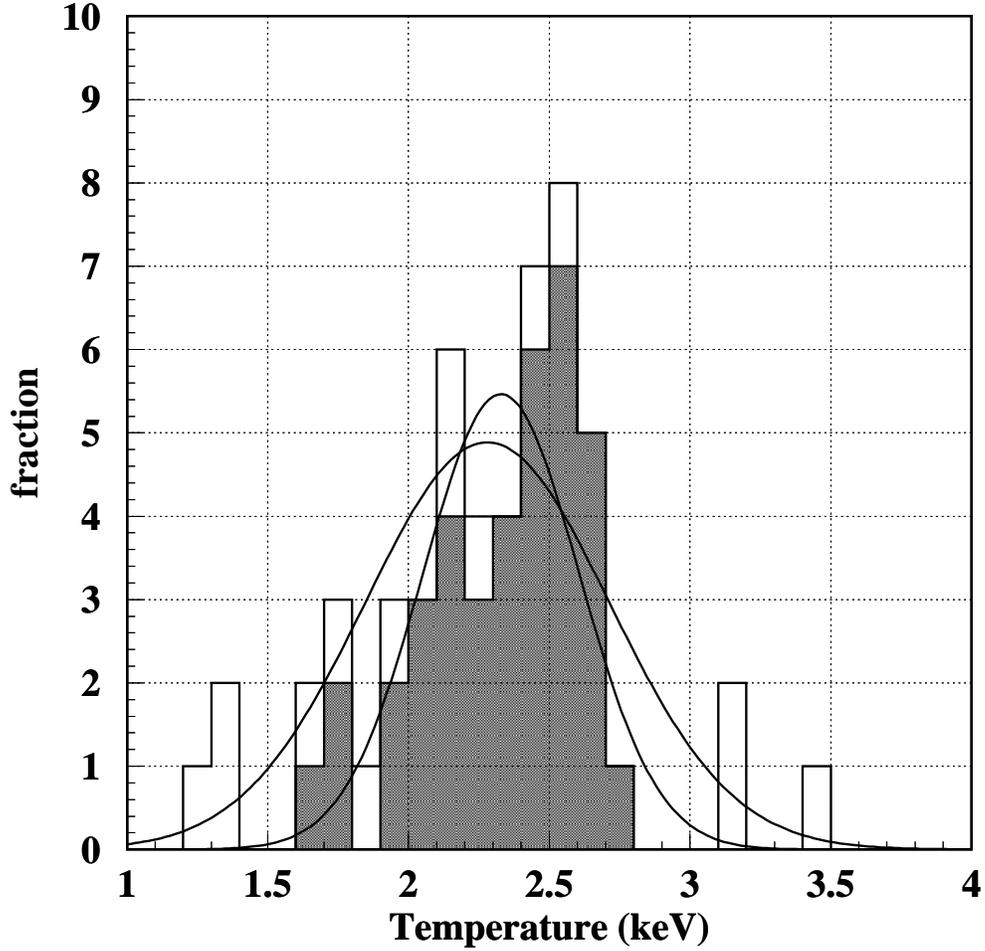,width=0.85\textwidth,angle=0}}
\caption{
Frequency distribution of the ICM temperatures based on the spectral
fits for the GIS FOV unit.  Shaded and unshaded portions indicate the
data within 2.5 $^{\circ}$ centered on M87 and the all data in the
Virgo field, respectively.  We fitted these distributions with a
Gaussian model.  Best-fit parameters are obtained to be an average
temperature of $\bar{T}$ = 2.33 keV with $\sigma_{\bar{T}}$ = 0.28 keV
within $2.5^\circ$ from M87 and $\bar{T}$ = 2.28 keV with
$\sigma_{\bar{T}}$ = 0.43 keV for all regions.
}
\end{center}
\end{figure}

\begin{figure}[h]
\begin{center}
\vspace*{5mm}
\mbox{\psfig{figure=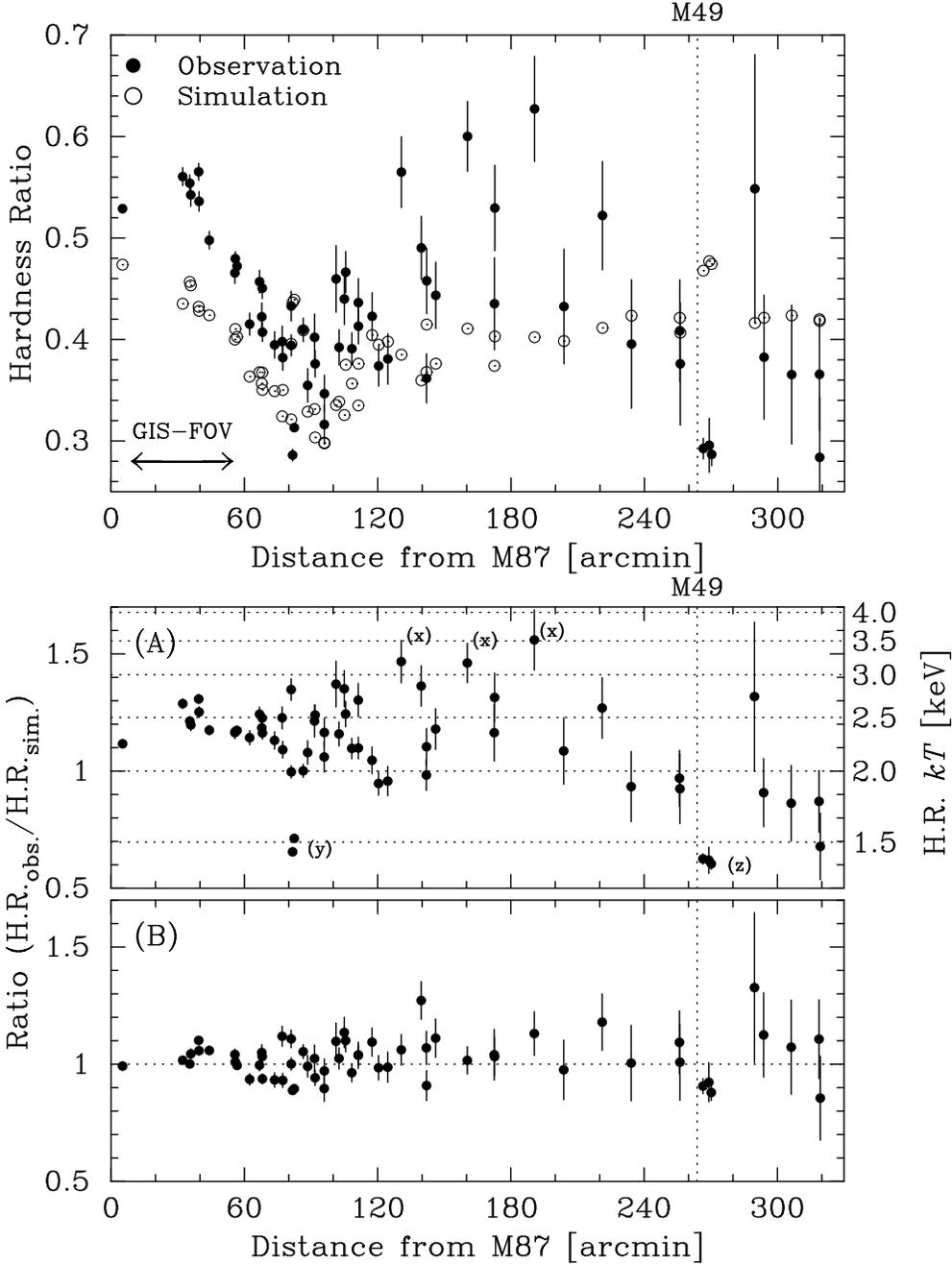,width=0.65\textwidth,angle=0}}
\vspace*{5mm}
\caption{
Hardness ratio from the counting rates between energy bands 2.0 $-$
7.0 keV and 0.7 $-$ 2.0 keV for each pointed field ($r \sim 22'$).
({\it top}): Hardness ratios from the raw counting rates (filled
circles, H.R.$_{\rm obs.}$) and from the ray-tracing simulation (open
circles, H.R.$_{\rm sim.}$), in which the MEKAL model with an uniform
temperature of 2.0 keV and a metal abundance of 0.3 solar are assumed.
({\it middle}): Corrected hardness ratio (= H.R.$_{\rm
obs.}$/H.R.$_{\rm sim.}$) which take account of the difference in the
response based on the simulated isothermal profile.  Notations of (x),
(y), and (z) indicate the high temperature region reported by Kikuchi
(2000), the regions which include M86 and M49, respectively.  ({\it
bottom}): Same as a middle panel, but showing the corrected hardness
ratio from the ray-tracing simulation, in which the temperature
distribution of the middle panel is assumed.  In all panels, errors
indicate 1 $\sigma$ statistical limits.
}
\end{center}
\end{figure}

\begin{figure}[h]
\begin{center}
\mbox{\psfig{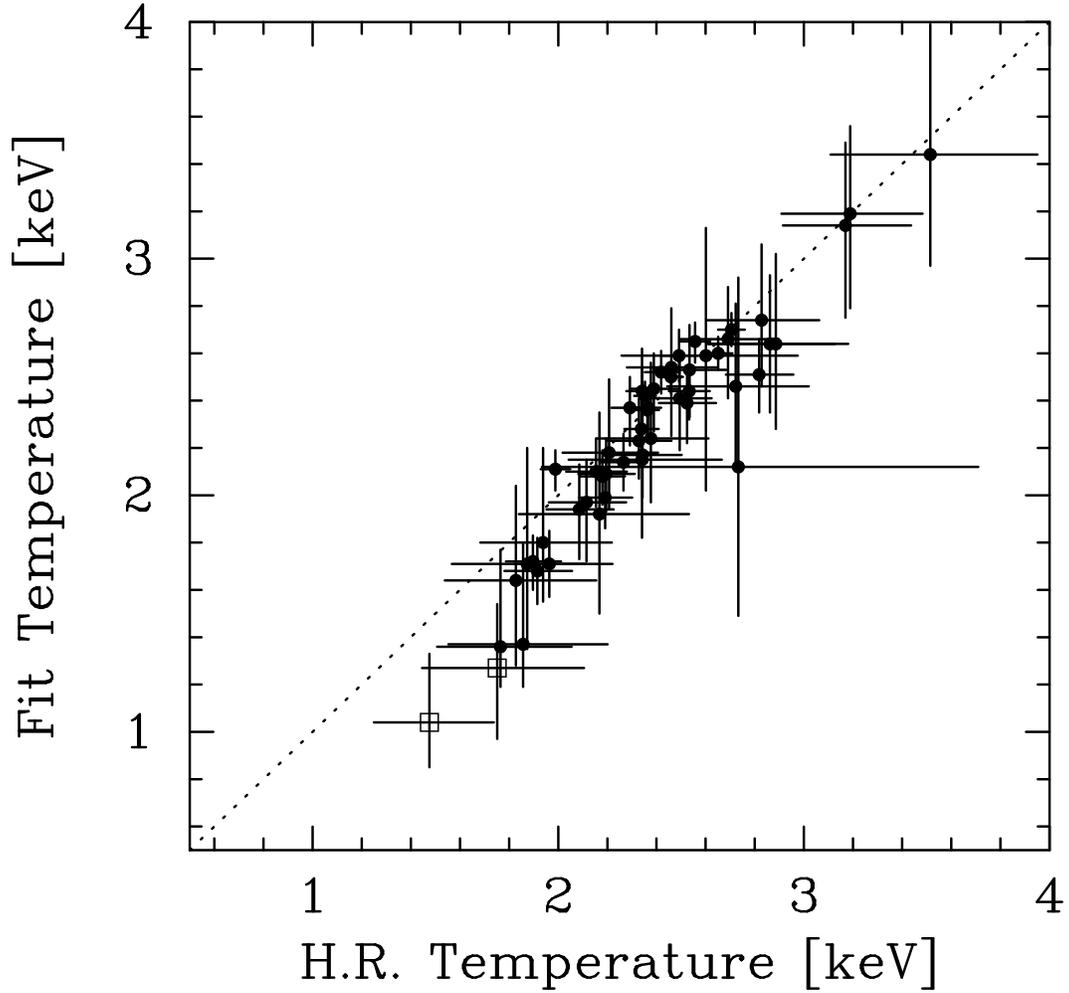}}
\caption{
Correlation between the temperatures derived from the hardness ratio
(H.R.) and from the spectral fit, except for some regions which
require two temperature components.  Dot line indicates the consistent
temperature.  Because two regions (pos-5002S and 5003S, open square)
indicate much lower metal abundance ($<$ 0.1 solar) and temperature
($<$ 1.3 keV) compared with the model spectrum for the simulation,
both temperatures are inconsistent.  If we fit this correlation with a
linear function, a slope is obtained to be 0.99 $\pm$ 0.01 with a
$\chi^2_{\nu}$ value of 0.72 (d.o.f. = 51).
}
\end{center}
\end{figure}

\begin{figure}[h]
\begin{center}
\vspace{-7mm}
\mbox{\psfig{figure=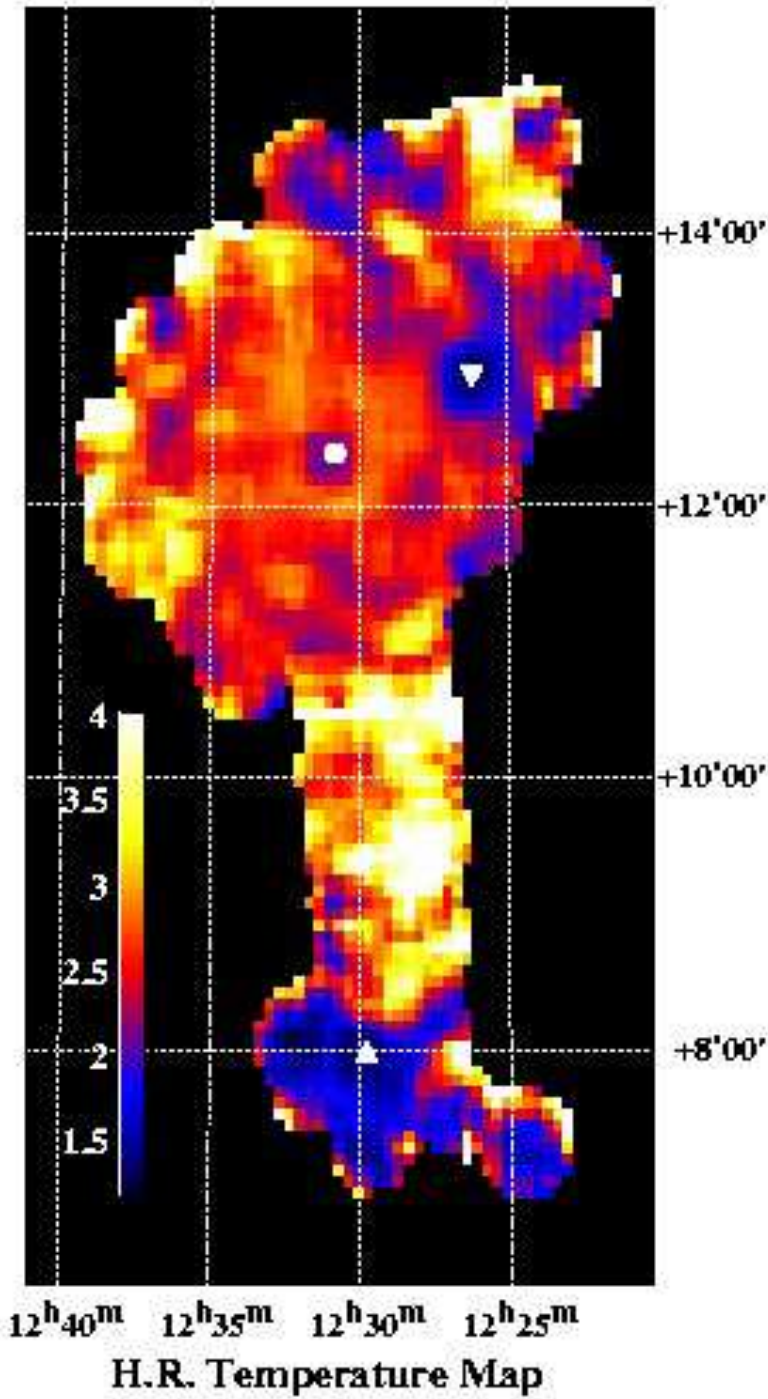,width=0.65\textwidth,angle=0}}
\vspace{-4mm}
\caption{
Temperature map of the Virgo cluster, based on the temperature
estimation from hardness ratio. The difference of the response for
each pointed field and the vignetting effect are taken into account.
Each pixel size is $5' \times 5'$ square, and H.R. temperature is
calculated from the average photon number ratio of the surrounding
$20' \times 20'$ square in order to improve photon statistics.  The
source mask mentioned in section 2.2 is applied.  Note
that the average temperatures between the ICM and ISM are shown at the
position of M87 (circle), M86 (inverse triangle) and M49 (triangle).
}
\end{center}
\end{figure}

\begin{figure}[h]
    \begin{minipage}{0.58\textwidth}
\begin{center}
\mbox{\psfig{figure=./Fig_9a.ps,width=0.9\textwidth,angle=270}}
\mbox{\psfig{figure=./Fig_9b.ps,width=0.9\textwidth,angle=270}}
\mbox{\psfig{figure=./Fig_9c.ps,width=0.9\textwidth,angle=270}}
\mbox{\psfig{figure=./Fig_9d.ps,width=0.9\textwidth,angle=270}}
\mbox{\psfig{figure=./Fig_9e.ps,width=0.9\textwidth,angle=270}}
\end{center}
    \end{minipage}
    \hfill
    \begin{minipage}{0.35\textwidth}
\hspace*{-18mm}
\mbox{\psfig{figure=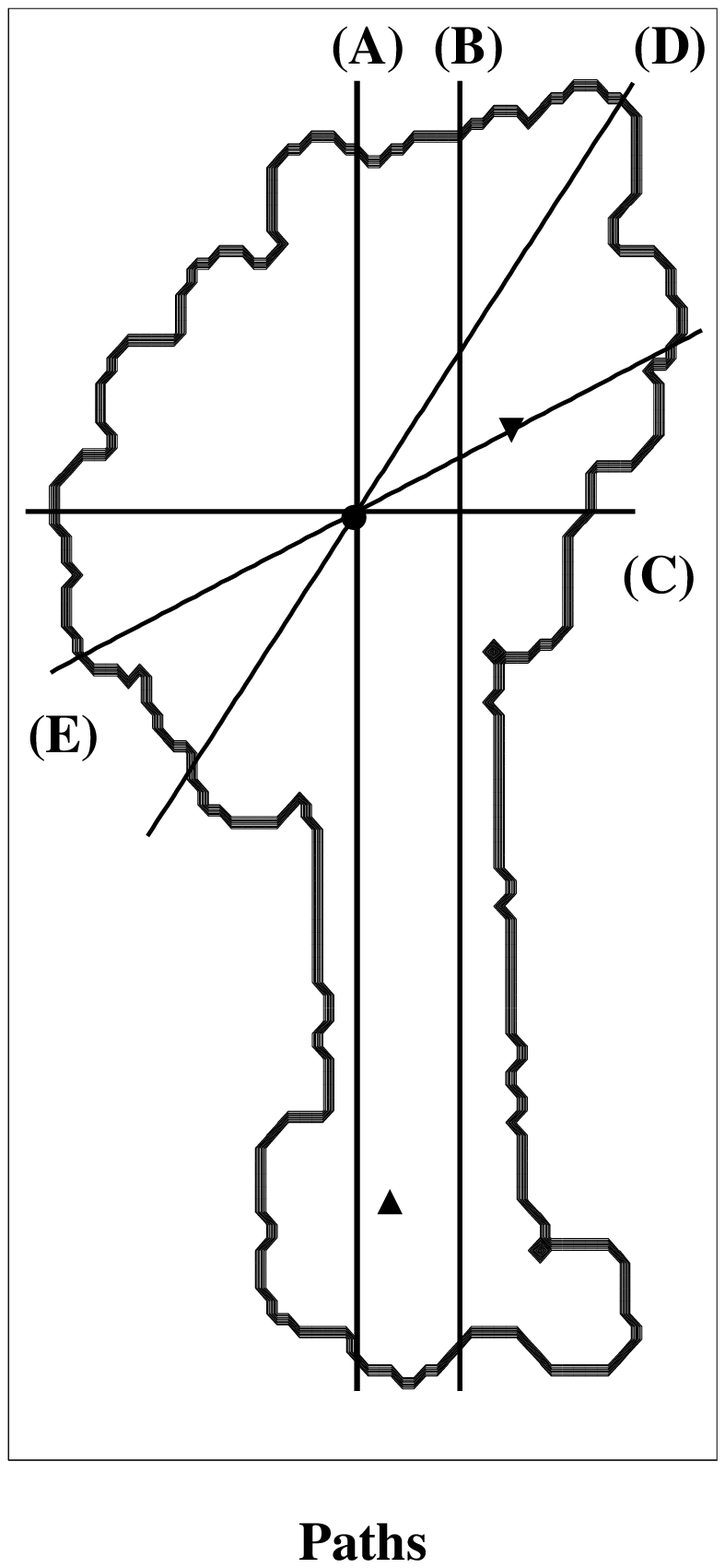,width=0.98\textwidth,angle=0}}\\
\caption{
({\it left}): Cross sections of the H.R. temperature map in figure
8.  The H.R. temperature (= corrected H.R.) in each
bin is calculated for a $20' \times 20'$ square.  Open squares show
the region including the ISM component of M87, M86 , and M49.  ({\it
right}): Calculated paths are indicated (see text).  The positions of
three galaxies are marked: M87 (circle), M86 (inverse triangle), M49
(triangle).
}
    \end{minipage}
\end{figure}

\begin{figure}[h]
\begin{center}
\mbox{\psfig{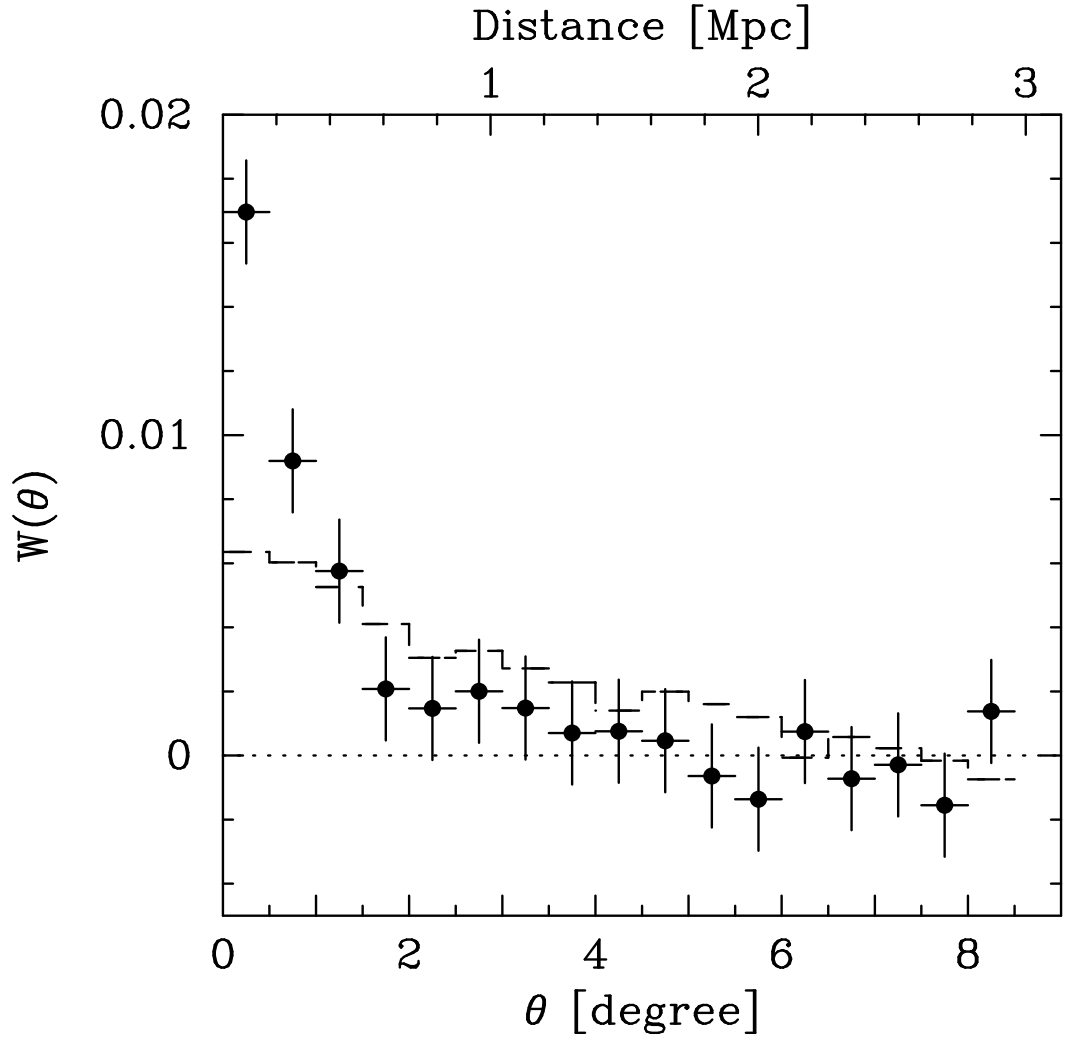}}
\caption{
Auto-correlation function (ACF) for the two dimensional C.H.R. based
on the observed data (filled circle) and the model data assuming the
large-scale temperature gradient centered on M87 (broken line).
}
\end{center}
\end{figure}

\begin{figure}[h]
\begin{center}
\mbox{\psfig{figure=./Fig_11a.ps,width=0.49\textwidth,angle=270}}
\mbox{\psfig{figure=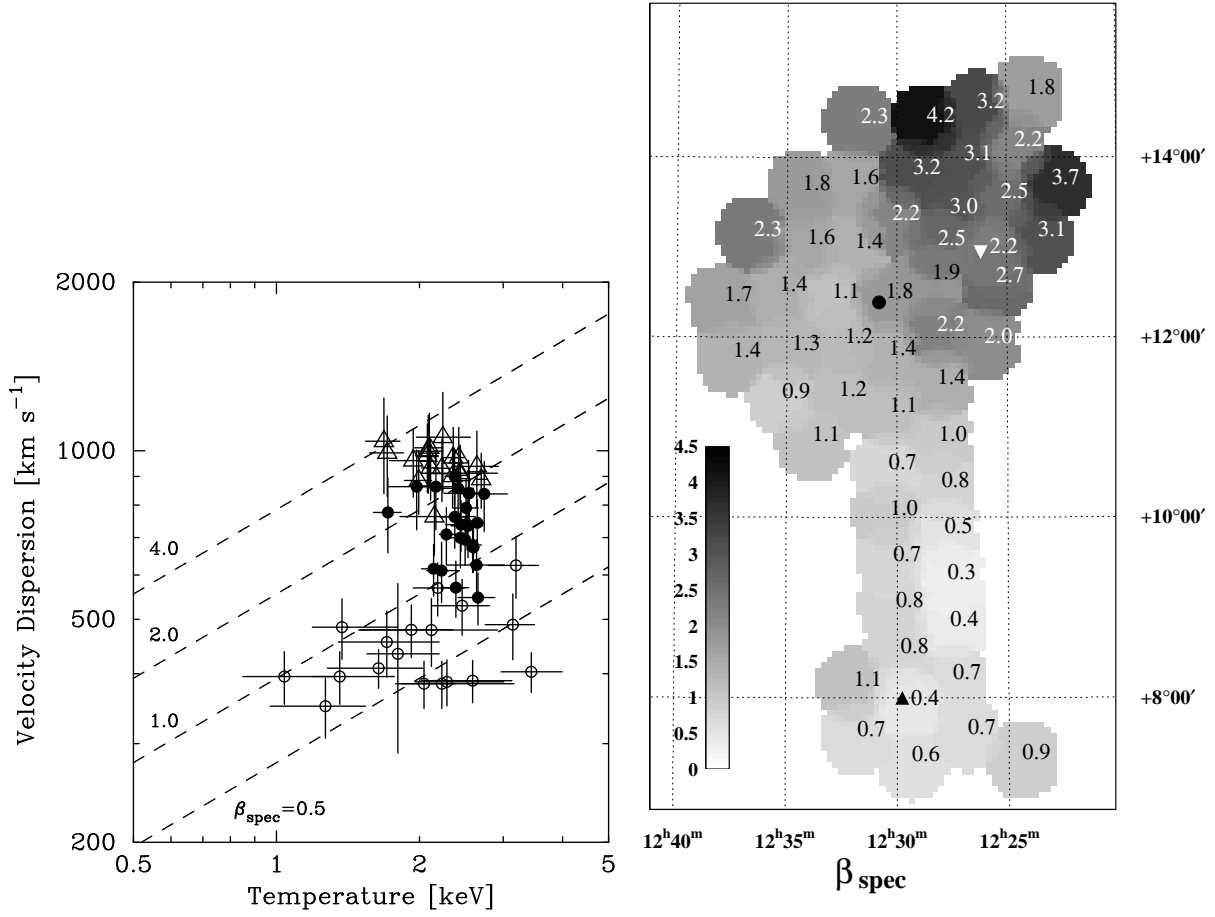,width=0.49\textwidth,angle=0}}
\caption{
({\it left}): Correlation between the velocity dispersions
($\sigma^2$) and the ICM temperatures ($kT$).  The velocity
dispersions are calculated from mean values within a radius of
1.5$^{\circ}$ centered on the GIS-FOV. The symbols indicate different
regions. Triangles are for the northwest of M87 (around M86), filled
circles are within $2.5^\circ$ from M87 except for the northwest
region, and open circles are the remaining regions, respectively.
Broken lines indicate $\beta_{\rm spec}$ (= $\mu m_p \sigma^2 / kT$)
values = 0.5, 1.0, 2.0 and 4.0, respectively.  ({\it right}):
$\beta_{\rm spec}$ map of the Virgo cluster.  The positions of
three bright galaxies are marked: M87 (circle), M86 (inverse
triangle), M49 (triangle).
}
\end{center}
\end{figure}

\begin{figure}[h]
\begin{center}
\vspace*{-3mm}
\mbox{\psfig{figure=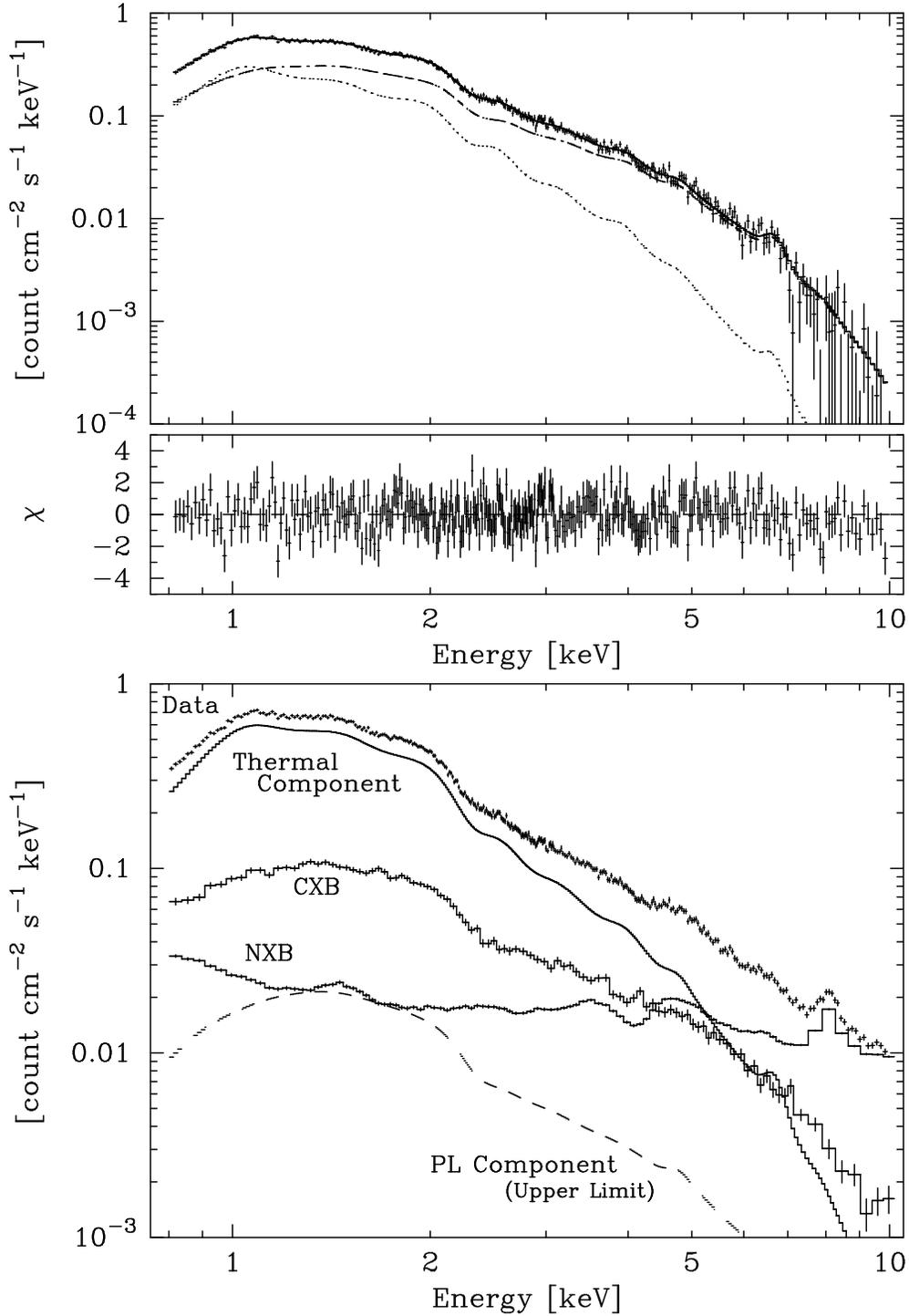,width=0.85\textwidth,angle=0}}
\vspace*{7mm}
\caption{
({\it top}): The Average Virgo spectrum within r = 0.4$^{\circ} -
2.5^{\circ}$ centered on M87 (cross mark), the best-fit two
temperature MEKAL model (line) which consists of hot and cool
components (broken and dot lines) , and the residual between the data
and model.  ({\it bottom}): Each component in the above spectrum
(observed data, CXB, and NXB spectra) together with the model spectrum
(thermal component and upper limit of the PL component).
}
\end{center}
\end{figure}

\clearpage

\begin{longtable}{lccrcrccl}
\caption{Log of ASCA  observations and result of spectral fits
} \\
\hline \hline  
Pointing$^{\dag}$ & \multicolumn{2}{c}{R.A., Dec.} & Dist.$^{\ddag}$ & Date$^{\S}$  & Exp.$^{*}$ & Rate$^{\P}$ 
&  Temperature~$^{\pounds}$ & $\chi^{2}$ (d.o.f)\\
Name  & \multicolumn{2}{c}{\small [deg.](J2000)}  & \footnotesize[arcmin.]& \small[UT]  & \small[sec]~~~ & \small[c/s] & 
\small [keV] & \\ 
\hline
&&&&&&&& \\[-3mm]
\endfirsthead
\multicolumn{9}{l}{\small\sl continued from previous page}\\
\hline  
Pointing$^{\dag}$ & \multicolumn{2}{c}{R.A., Dec.} & Dist.$^{\ddag}$ & Date$^{\S}$  & Exp.$^{*}$ & Rate$^{\P}$ 
&  Temperature~$^{\pounds}$ & $\chi^{2}$ (d.o.f) \\
Name  & \multicolumn{2}{c}{\small [deg.](J2000)}  & \footnotesize[arcmin.]& \small[UT]  & \small[sec]~~~ & \small[c/s] & 
\small [keV] & \\ 
\hline
&&&&&&&& \\[-3mm]
\endhead
\hline\multicolumn{9}{r}{\small\sl continued on next page}\\
\hline
\endfoot
&&&&&&&& \\[-3mm]
\hline
\hline
\multicolumn{9}{p{16cm}}{\small $\dag$: $ddnnD$ ($dd$, $nn$, and $D$ show the 
distance range between $d.d^{\circ}$ and $d.d^{\circ} + 0.5^{\circ}$ from M87 
, serial number, and azimuth direction from M87 , i.e., ``North, West, 
South and East'', respectively).} \\
\multicolumn{9}{p{16cm}}{\small $\ddag$: Distance from the center of M87 in arcminutes.	
The position of M87 is (R.A, Dec)$_{\rm J2000}$ = (187.7058, 12.3911).} \\
\multicolumn{9}{p{16cm}}{\small $\S$: Start time of observation in year/month/day.} \\
\multicolumn{9}{p{16cm}}{\small $*$: Total Exposure time of GIS (per sensor) after the data screening.} \\
\multicolumn{9}{p{16cm}}{\small $\P$: GIS counting rate (per sensor, 0.7-10.0 keV band) including 
the cosmic X-ray background and non X-ray background.} \\
\multicolumn{9}{p{16cm}}{\small $\pounds$: 
ICM temperature estimated by spectral fit with the MEKAL model (see section 
3.1).} \\
\multicolumn{9}{p{16cm}}{\small (a)(b)(c):
Pointings including three bright galaxies, M87, M86 and M49, respectively.
Two temperature MEKAL model for the ICM and ISM was adopted.
} \\
\endlastfoot
 0001$~^{\small (a)}$ & 187.63 & 12.43 &   5 ~ & \small 93/06/07  & 12017  & 8.278 & 2.54$^{+0.12}_{-0.06}$, 1.12$^{+0.16}_{-0.07}$ & 916.2 (537) \\
&&&&&&&& \\[-3mm]				 		                                                  	     
0501N & 187.55 & 13.31 &  56 ~ & \small 97/06/14  &  9461  & 1.015 & 2.44$^{+0.10}_{-0.11}$               & 105.1 (95) \\
0502N & 187.03 & 13.03 &  56 ~ & \small 95/12/24  & 20784  & 1.079 & 2.42$^{+0.06}_{-0.07}$               & 147.0 (95) \\
0503N & 187.02 & 13.04 &  57 ~ & \small 96/01/05  & 17820  & 1.092 & 2.36$^{+0.07}_{-0.13}$               & 176.0 (95) \\
0504W & 187.10 & 12.65 &  39 ~ & \small 93/06/08  & 11392  & 1.342 & 2.70$^{+0.07}_{-0.07}$               & 161.7 (95) \\
0505W & 187.04 & 12.07 &  44 ~ & \small 99/06/21  & 10159  & 1.240 & 2.37$^{+0.08}_{-0.10}$               & 142.9 (95) \\
0506S & 187.59 & 11.80 &  36 ~ & \small 98/06/11  &  4811  & 1.752 & 2.52$^{+0.09}_{-0.09}$             & 105.4 (95) \\
0507S & 188.09 & 11.94 &  35 ~ & \small 99/06/23  &  9227  & 1.584 & 2.50$^{+0.06}_{-0.07}$             & 124.3 (95) \\
0508E & 188.24 & 12.44 &  32 ~ & \small 99/06/23  &  8596  & 1.501 & 2.60$^{+0.07}_{-0.07}$             & 120.1 (95) \\
0509N & 187.97 & 12.99 &  40 ~ & \small 99/06/23  & 10486  & 1.136 & 2.65$^{+0.08}_{-0.09}$             & 116.2 (95) \\
&&&&&&&& \\[-3mm]				 		                                      	     
1001N & 187.32 & 13.81 &  89 ~ & \small 96/12/27  &  7902  & 0.600 & 2.10$^{+0.14}_{-0.15}$             & 82.3 (64) \\
1002N & 186.89 & 13.39 &  77 ~ & \small 96/12/27  &  8308  & 0.786 & 2.08$^{+0.11}_{-0.12}$             & 74.0 (64) \\
1003W$~^{\small (b)}$ & 186.47 & 12.96 &  82 ~ & \small 93/07/03  & 16920  & 1.019 & 2.17$^{+0.14}_{-0.13}$, 0.78 (fix)  & 128.4 (94)      \\
1004W$~^{\small (b)}$ & 186.44 & 12.93 &  82 ~ & \small 98/06/09  & 76313  & 1.036 & 2.35$^{+0.07}_{-0.09}$, 0.78 (fix)  & 172.4 (94)      \\
1005W$~^{\small (b)}$ & 186.33 & 12.81 &  86 ~ & \small 93/07/04  & 18314  & 0.858 & 2.35$^{+0.26}_{-0.26}$, 0.78 (fix)  & 79.0 (70)    \\
1006W & 186.37 & 12.61 &  81 ~ & \small 95/06/21  & 20218  & 0.732 & 2.11$^{+0.08}_{-0.09}$             & 88.3 (62) \\
1007W & 186.50 & 11.94 &  77 ~ & \small 99/06/22  & 10434  & 0.576 & 2.41$^{+0.19}_{-0.22}$             & 89.1 (64) \\
1008S & 187.04 & 11.48 &  68 ~ & \small 98/06/11  & 10323  & 0.669 & 2.45$^{+0.15}_{-0.18}$             & 87.2 (64) \\
1009S & 187.58 & 11.17 &  74 ~ & \small 98/06/11  & 12379  & 0.651 & 2.14$^{+0.12}_{-0.12}$             & 78.0 (64) \\
1010S & 188.15 & 11.35 &  68 ~ & \small 99/06/22  & 12975  & 0.827 & 2.59$^{+0.11}_{-0.11}$             & 72.1 (64) \\
1011E & 188.69 & 11.86 &  67 ~ & \small 99/06/30  & 12530  & 0.762 & 2.44$^{+0.11}_{-0.12}$             & 79.5 (64) \\
1012E & 188.83 & 12.52 &  68 ~ & \small 95/06/26  & 18392  & 0.749 & 2.28$^{+0.12}_{-0.08}$             & 76.9 (64) \\
1013E & 188.52 & 13.04 &  62 ~ & \small 99/07/02  & 10825  & 0.788 & 2.37$^{+0.13}_{-0.16}$             & 64.1 (64) \\
1014N & 188.02 & 13.71 &  81 ~ & \small 99/06/24  & 15105  & 0.531 & 2.51$^{+0.15}_{-0.16}$             & 72.1 (64) \\
&&&&&&&& \\[-3mm]				 		                                      	     
1501N & 186.73 & 13.97 & 111 ~ & \small 97/06/16  & 17547  & 0.473 & 2.09$^{+0.15}_{-0.16}$             & 85.6 (64) \\
1502W & 186.31 & 13.54 & 108 ~ & \small 96/12/26  & 19271  & 0.445 & 1.99$^{+0.13}_{-0.13}$             & 77.1 (64) \\
1503W & 185.89 & 13.12 & 117 ~ & \small 96/12/26  & 17043  & 0.384 & 1.94$^{+0.19}_{-0.21}$             & 85.4 (64) \\
1504S & 187.03 & 10.85 & 101 ~ & \small 98/06/12  & 10210  & 0.364 & 2.64$^{+0.38}_{-0.36}$             & 72.3 (64) \\
1505S & 187.59 & 10.54 & 111 ~ & \small 96/06/12  & 31134  & 0.522 & 2.66$^{+0.22}_{-0.25}$             & 69.1 (64) \\
1506S & 188.46 & 10.85 & 103 ~ & \small 99/06/26  & 15839  & 0.442 & 2.23$^{+0.18}_{-0.16}$             & 71.2 (64) \\
1507E & 188.80 & 11.32 &  92 ~ & \small 99/06/26  & 19172  & 0.519 & 2.39$^{+0.15}_{-0.17}$             & 88.7 (64) \\
1508E & 189.36 & 11.78 & 106 ~ & \small 95/06/25  & 22836  & 0.485 & 2.53$^{+0.19}_{-0.20}$              & 75.0 (64) \\
1509E & 189.46 & 12.40 & 105 ~ & \small 99/07/01  & 16227  & 0.366 & 2.74$^{+0.32}_{-0.28}$             & 81.4 (64) \\
1510E & 189.13 & 13.12 &  96 ~ & \small 97/06/24  & 16054  & 0.432 & 2.17$^{+0.18}_{-0.18}$             & 75.0 (64) \\
1511E & 189.13 & 13.13 &  96 ~ & \small 97/07/06  & 14472  & 0.427 & 1.97$^{+0.18}_{-0.25}$             & 80.5 (64) \\
1512N & 188.57 & 13.65 &  92 ~ & \small 99/06/24  & 10455  & 0.432 & 2.54$^{+0.25}_{-0.29}$             & 60.8 (64) \\
1513N & 187.92 & 14.39 & 120 ~ & \small 97/06/22  & 28349  & 0.397 & 1.72$^{+0.11}_{-0.12}$             & 93.1 (64) \\
&&&&&&&& \\[-3mm]				 		                                      	     
2001N & 187.16 & 14.39 & 125 ~ & \small 97/06/17  & 16349  & 0.356 & 1.68$^{+0.14}_{-0.14}$             & 73.6 (64) \\
2002N & 186.58 & 14.55 & 146 ~ & \small 97/06/16  & 16466  & 0.336 & 2.24$^{+0.32}_{-0.27}$             & 45.6 (64) \\
2003N & 186.15 & 14.12 & 140 ~ & \small 97/06/15  & 17252  & 0.351 & 2.64$^{+0.29}_{-0.29}$             & 78.3 (64) \\
2004W & 185.73 & 13.70 & 142 ~ & \small 96/12/25  & 18667  & 0.355 & 1.71$^{+0.14}_{-0.14}$             & 98.7 (64) \\
2005S & 187.00 & 10.33 & 131 ~ & \small 98/06/12  & 19671  & 0.334 & 3.19$^{+0.37}_{-0.40}$              & 82.5 (64) \\
2006S & 187.58 & 10.03 & 142 ~ & \small 98/06/12  & 14419  & 0.346 & 2.18$^{+0.31}_{-0.24}$             & 82.1 (64) \\
&&&&&&&& \\[-3mm]				 		                                      	     
2501N & 186.00 & 14.71 & 172 ~ & \small 97/06/14  & 22875  & 0.281 & 2.15$^{+0.47}_{-0.33}$             & 59.8 (64) \\
2502S & 186.96 &  9.82 & 160 ~ & \small 98/06/13  & 19100  & 0.359 & 3.14$^{+0.35}_{-0.39}$             & 104.1 (64) \\
2503S & 187.55 &  9.52 & 173 ~ & \small 98/06/13  & 17033  & 0.299 & 2.46$^{+0.35}_{-0.33}$             & 62.1 (64) \\
&&&&&&&& \\[-3mm]				 		                                      	     
3001S & 186.93 &  9.31 & 191 ~ & \small 98/06/17  & 18154  & 0.294 & 3.44$^{+0.55}_{-0.47}$             & 61.8 (64) \\
3002S & 187.51 &  9.00 & 204 ~ & \small 98/06/17  & 17335  & 0.264 & 1.92$^{+0.43}_{-0.42}$             & 62.0 (64) \\
&&&&&&&& \\[-3mm]				 		                                      	     
3501S & 186.90 &  8.80 & 221 ~ & \small 98/06/18  & 16544  & 0.402 & 2.59$^{+0.54}_{-0.57}$             & 91.7 (64) \\
3502S & 187.46 &  8.50 & 234 ~ & \small 98/06/23  & 10070  & 0.285 & 1.71$^{+0.49}_{-0.36}$             & 57.4 (64) \\
&&&&&&&& \\[-3mm]				 		                                      	     
4001S & 186.87 &  8.21 & 256 ~ & \small 98/06/18  & 14268  & 0.284 & 1.80$^{+0.40}_{-0.25}$              & 84.0 (64) \\
4002S$~^{\small (c)}$ & 187.41 &  7.96 & 266 ~ & \small 93/07/04  & 17256  & 0.547 & 2.28$^{+0.73}_{-0.32}$, 0.81$^{+0.07}_{-0.10}$  & 144.9 (93) \\
4003S & 187.98 &  8.13 & 256 ~ & \small 98/06/20  & 14092  & 0.271 & 1.37$^{+0.43}_{-0.18}$              & 73.6 (64) \\
&&&&&&&& \\[-3mm]				 		                                                  	     
4501S & 186.71 &  7.60 & 294 ~ & \small 98/06/19  & 10811  & 0.269 & 1.64$^{+0.40}_{-0.36}$               & 82.8 (64) \\
4502S$~^{\small (c)}$ & 187.34 &  7.92 & 269 ~ & \small 93/06/30  &  2718  & 0.537 & 2.23$^{+0.93}_{-0.45}$, 0.73$^{+0.14}_{-0.11}$ & 98.7 (93) \\
4503S$~^{\small (c)}$ & 187.35 &  7.90 & 270 ~ & \small 93/06/30  & 16977  & 0.515 & 2.04$^{+0.34}_{-0.21}$, 0.82$^{+0.08}_{-0.11}$ & 119.0 (93) \\
4504S & 187.94 &  7.57 & 290 ~ & \small 98/06/19  &  4476  & 0.262 & 2.12$^{+0.80}_{-0.63}$                & 69.5 (64) \\
&&&&&&&& \\[-3mm]				 		                                                 	     
5001S & 186.10 &  7.33 & 319 ~ & \small 93/06/27  & 18877  & 0.256 & 1.36$^{+0.41}_{-0.17}$               & 70.2 (64) \\
5002S & 186.10 &  7.33 & 319 ~ & \small 93/06/28  & 11799  & 0.259 & 1.04$^{+0.29}_{-0.19}$              & 79.9 (64) \\
5003S & 187.34 &  7.30 & 306 ~ & \small 98/12/21  & 11764  & 0.259 & 1.27$^{+0.27}_{-0.30}$               & 67.9 (64) \\
&&&&&&&& \\[-3mm]
\end{longtable}

\begin{table}
\begin{center}
\caption{Best-fit parameters of the whole Virgo spectrum}
\begin{tabular}{lcc} \hline \hline 
 &  MEKAL model &~~ MEKAL + MEKAL model \\ \hline
$\chi^2$ (d.o.f.) & 395.7 (322) & 351.4 (320) \\
$kT$ ~[keV]& 2.50 $^{+0.04}_{-0.05}$ 
& 3.67 $^{+0.53}_{-0.52}$, 1.38 $^{+0.09}_{-0.13}$ \\
$Z_{\rm Fe}$ ~[solar] & 0.26 $^{+0.04}_{-0.04}$ & 0.16 $^{+0.04}_{-0.03}$ \\
$Z_{\rm S}$  ~[solar] & 0.50 $^{+0.12}_{-0.11}$ & 0.52 $^{+0.10}_{-0.09}$ \\
$Z_{\rm Si}$ ~[solar] & 0.45 $^{+0.11}_{-0.10}$ & 0.33 $^{+0.08}_{-0.06}$ \\
$F_X^{\rm thermal}$(2-10keV) ~{\small [ergs cm$^{-2}$ s$^{-1}$ arcmin$^{-2}$]}
~~ & (1.05 $^{+0.07}_{-0.07}) \times 10^{-14}$ &
(1.09 $^{+0.23}_{-0.14}) \times 10^{-14}$ \\
&& \\[-3mm] \hline
 &   & + Power-Law model \\ \hline
$\chi^2$ (d.o.f.) &  & 351.4 (319) \\
Photon Index & & 1.7 (fixed) \\
$F_X^{\rm hard}$(2-10keV) ~{\small [ergs cm$^{-2}$ s$^{-1}$ arcmin$^{-2}$]}
 & &  $< 9.52 \times 10^{-16}$ \\ 
$L_X^{\rm hard}$(2-10keV) ~[ergs s$^{-1}$] & & 
$< 2.28 \times 10^{42}$ \\ 
Flux Ratio (= $F_X^{\rm hard}$/$F_X^{\rm thermal}$) & &
$< 0.09 $ \\ \hline \hline 
%
\end{tabular}
\end{center}
\end{table}

\end{document}